\documentclass[twocolumn]{aastex631}

\usepackage{multirow}

\usepackage[utf8]{inputenc}
\usepackage{natbib}
\usepackage{graphicx}
\usepackage{longtable}
\usepackage{enumerate}
\usepackage{amsmath}
\usepackage{hyperref}
\usepackage{CJK}
\usepackage[title]{appendix}
\usepackage{rotating}
\usepackage{longtable}

\usepackage{amsmath}

\def\msun{\ifmmode {\rm M}_{\mathord\odot}\else $M_{\mathord\odot}$\fi}
\def\rsun{\ifmmode {\rm R}_{\mathord\odot}\else $R_{\mathord\odot}$\fi}
\def\lsun{\ifmmode {\rm L}_{\mathord\odot}\else $L_{\mathord\odot}$\fi}

\def\c18o{C$^{18}$O}
\def\h2{H$_{2}$}
\def\13co{$^{13}$CO}
\def\n2hp{$_{2}$H$^{+}$}

\def\cm2{cm$^{-2}$}

\newcommand{\kms}{km~s$^{-1}$}

\def\orion{{\sc orion2}}

\shorttitle{}
\shortauthors{}

\begin{document}
\begin{CJK*}{UTF8}{gbsn}

\title{Exploring Magnetic Fields in Molecular Clouds through Denoising Diffusion Probabilistic Models}

\author[0000-0001-6216-8931]{Duo Xu}
\affiliation{Department of Astronomy, University of Virginia, Charlottesville, VA 22904-4235, USA}
\affiliation{Canadian Institute for Theoretical Astrophysics, University of Toronto, 60 St. George Street, Toronto, ON M5S 3H8, Canada}

\author{Jenna Karcheski}
\affiliation{University of Wisconsin-Madison, Department of Astronomy, 475 N Charter St, Madison, WI 53703, USA}

\author[0000-0003-1964-970X]{Chi-Yan Law}
\affiliation{Department of Space, Earth \& Environment, Chalmers University of Technology, SE-412 96 Gothenburg, Sweden}
\affiliation{European Southern Observatory, Karl-Schwarzschild-Strasse 2, D-85748 Garching, Germany}

\author[0000-0003-0774-9375]{Ye Zhu}
\affiliation{Department of Computer Science, Princeton University, Princeton NJ 08544, USA}

\author{Chia-Jung Hsu}
\affiliation{Department of Space, Earth \& Environment, Chalmers University of Technology, SE-412 96 Gothenburg, Sweden}

\author[0000-0002-3389-9142]{Jonathan C. Tan}
\affiliation{Department of Astronomy, University of Virginia, Charlottesville, VA 22904-4235, USA}
\affiliation{Department of Space, Earth \& Environment, Chalmers University of Technology, SE-412 96 Gothenburg, Sweden}

\email{xuduo@cita.utoronto.ca}

\begin{abstract}

Accurately measuring magnetic field strength in the interstellar medium, including giant molecular clouds (GMCs), remains a significant challenge. We present a machine learning approach using Denoising Diffusion Probabilistic Models (DDPMs) to estimate magnetic field strength from synthetic observables such as column density, dust continuum polarization vector orientation angles, and line-of-sight (LOS) nonthermal velocity dispersion. We trained three versions of the DDPM model: the 1-channel DDPM (using only column density), the 2-channel DDPM (incorporating both column density and polarization angles), and the 3-channel DDPM (which combines column density, polarization angles, and LOS nonthermal velocity dispersion). We assessed the models on both synthetic test samples and new simulation data that were outside the training set's distribution. The 3-channel DDPM consistently outperformed both the other DDPM variants and the power-law fitting approach based on column density alone, demonstrating its robustness in handling previously unseen data. Additionally, we compared the performance of the Davis-Chandrasekhar-Fermi (DCF) methods, both classical and modified, to the DDPM predictions. The classical DCF method overestimated the magnetic field strength by approximately an order of magnitude. Although the modified DCF method showed improvement over the classical version, it still fell short of the precision achieved by the 3-channel DDPM.
\end{abstract}

\section{Introduction}
\label{Introduction}

Magnetic fields are a ubiquitous and significant element of galactic environments, permeating the interstellar medium (ISM) \citep[e.g.,][]{1999ApJ...520..706C,2017ARA&A..55..111H}. They play a crucial role in numerous astrophysical processes and have a profound influence on the structure, dynamics, and evolution of the ISM \citep{2012ARA&A..50...29C,2015MNRAS.450.4035F}. Magnetic fields interact with the gas and dust in the ISM, exerting pressure on the gas and affecting its dynamics, including providing support against gravitational collapse. On large scales, magnetic fields can exhibit coherent structures such as spiral arms and filamentary structures spanning hundreds of parsecs {\citep[e.g.,][]{1999A&A...348..405H,2024A&A...686L..11W}.} However, on smaller scales, the magnetic field becomes more complex, entangled, and turbulent, shaped by the interplay between gas dynamics and stellar feedback {\citep[e.g.,][]{2021ApJ...912L..27E,2023ApJ...952...29K}.} In spite of the potential importance of magnetic fields in the ISM, accurately measuring their strength is a challenging task.

Observations of magnetic fields in the ISM can be categorized into two main types. The first type involves measurements of the plane-of-sky (POS) component, which is often traced using techniques such as polarized thermal dust emission \citep{1998ApJ...502L..75R,2016A&A...586A.138P}, starlight polarization \citep{1951ApJ...114..206D,2002ApJ...564..762F}, and synchrotron emission \citep{1982A&A...105..192B,2012ApJ...761L..11J}. The second type involves measurements of the line-of-sight (LOS) component, which is typically probed through Zeeman splitting \citep{1986ApJ...301..339T,2010ApJ...725..466C} and Faraday rotation \citep{1966MNRAS.133...67B,2022A&A...657A..43H}. However, accurately quantifying the total strength of magnetic fields in diverse ISM environments with these methods is difficult.

The strength of magnetic fields in the ISM shows significant variation across different regions {\citep{2017ARA&A..55..111H}}. On average, the magnetic field strength in the Milky Way is estimated to be around a few microgauss ($\rm \mu G$), but it ranges from fractions of a $\rm \mu G$ in diffuse areas {\citep{1999ApJ...520..706C,2010ApJ...725..466C,2022Natur.601...49C}} to several tens to thousands of $\rm \mu G$ in dense molecular clouds and star-forming regions {\citep{1999ApJ...520..706C,1999ApJ...514L.121C,2010ApJ...725..466C,2016A&A...591A..19P}.} Direct measurements of the LOS component of the magnetic field, $B_z$, are typically obtained through the Zeeman effect, which is based on the splitting of spectral lines in the presence of a magnetic field \citep{1986ApJ...301..339T,2010ApJ...725..466C}. Meanwhile, indirect measurements of the POS component rely on the Davis-Chandrasekhar-Fermi (DCF) method \citep{PhysRev.81.890.2,1953ApJ...118..113C,2015A&ARv..24....4B}, which often uses polarized thermal dust emission \citep{1998ApJ...502L..75R,2016A&A...586A.138P}. The DCF method assumes an equipartition between magnetic field energy and the turbulent kinetic energy of the gas and relates magnetic field strength to the polarization angle and other observable properties. Specifically, the relation between the gas density $\rho$, the nonthermal velocity dispersion $\sigma_{V}$, and the polarization angle dispersion $\sigma_{PA}$ gives the POS magnetic field strength $B_{POS}$ as:
\begin{equation} \label{eqn_dcf}
B_{POS}=f\sqrt{4\pi\rho}\frac{\sigma_{V}}{\sigma_{PA}},
\end{equation}
where $f$ is a correction factor. Recently, \citet{2021A&A...647A.186S} introduced a modified DCF method that accounts for compressible modes, expressed as:
\begin{equation}
\label{eqn_modified_dcf}
B_{POS}=\sqrt{2 \pi \rho} \frac{\sigma_{V}}{\sqrt{\sigma_{PA}}}.
\end{equation} 
However, there is significant uncertainty in measuring magnetic field strength using the DCF method, especially due to challenges in accurately determining angular dispersion. This can be influenced by contributions from ordered magnetic fields \citep{2009ApJ...696..567H,2017ApJ...846..122P}, line-of-sight averaging \citep{1990ApJ...362..545Z,1991ApJ...373..509M}, beam-smoothing \citep{2009ApJ...706.1504H}, and other observational effects \citep{2016ApJ...820...38H}. As angular dispersion is a statistical quantity, consistently estimating magnetic field strength at a pixel level across entire maps is difficult, with local fluctuations potentially affecting the results \citep{2021ApJ...910..161Y}. {Additionally, the substantial uncertainty in estimating the nonthermal velocity dispersion (i.e., the linewidth measurement) poses significant challenges to accurately determining the magnetic field strength. The LOS gas velocity, which can be used to estimate the gas volume density-a crucial parameter in the DCF method-adds to this complexity \citep{2022MNRAS.514.1575C}. More fundamentally, the anisotropic nature of MHD turbulence undermines a core assumption of the DCF method, introducing substantial uncertainties in magnetic field strength estimation \citep{2022ApJ...935...77L}. Furthermore, the presence of gravitational forces also violates the basic assumptions of the DCF method, contributing to its high level of uncertainty \citep{2022ApJ...925...30L}.  }


Recent advances in deep learning offer a promising alternative for connecting observable quantities to intrinsic physical properties, such as magnetic field strength. For instance, \citet{2020ApJ...905..172X,2020ApJ...890...64X} demonstrated that the Convolutional Approach to Structure Identification (CASI), based on convolutional neural networks (CNNs), can effectively separate gas associated with stellar feedback (e.g., stellar winds and outflows) from ambient clouds using molecular line emission. {Additionally, \citet{2023ApJ...942...95X} demonstrated that CNNs can effectively predict magnetic field directions based on gas morphology, surpassing simpler techniques such as gradient geometry approaches, particularly in sub-Alfv\'enic and trans-Alfv\'enic cloud environments \citep[e.g.,][]{2013ApJ...774..128S,2018ApJ...865...46L,2020MNRAS.496.4546H}.} {\citet{2019ApJ...882L..12P} demonstrated that CNNs can effectively extract detailed information from gas morphology, enabling accurate differentiation between varying levels of magnetization. This supports the potential of machine learning as a valuable tool for inferring magnetic field strength from observational data. Furthermore, \citet{2024MNRAS.52711240H} showed that CNNs can retrieve 3D magnetic field strength information from synthetic CO data with relatively high accuracy, even though the study covered a narrower range of magnetic field strengths. These findings further underscore the promising capabilities of machine learning in this area.} 

Recently, Denoising Diffusion Probabilistic Models (DDPMs) have emerged as powerful tools for image generation \citep{pmlr-v37-sohl-dickstein15,NEURIPS2020_diffusion} and are showing great potential for prediction tasks in astronomy. Inspired by thermodynamic principles, DDPMs have demonstrated their ability to generate realistic galaxy images \citep{2022MNRAS.511.1808S} and enhance interferometric image quality by reducing noise \citep{2023arXiv230509121W}. Furthermore, DDPMs have been successfully applied to segmentation tasks, such as identifying filamentary structures in dust emission maps \citep{2023ApJ...955..113X}. They have also been used to infer the volume density of the ISM from column density maps \citep{2023ApJ...950..146X} and estimate the interstellar radiation field strength from multi-band dust emission \citep{2023ApJ...958...97X}. \citet{2023ApJ...950..146X,2023ApJ...958...97X} demonstrate the adaptability of DDPMs in linking observable data to intrinsic physical properties, even when faced with previously unseen data. This suggests that DDPMs could potentially surpass the DCF method in estimating magnetic field strength, particularly by accounting for deviations from the DCF assumptions—such as non-isotropic turbulence, incorrect angular dispersion tracing, or the lack of energy equipartition—through their ability to correct domain shifts during prediction. 

In this paper, we present a deep learning approach using denoising diffusion probabilistic models to estimate the magnetic field strength of GMCs from column density maps, dust polarization angles, and LOS velocity dispersions. In Section~\ref{Data and Method}, we explain the diffusion model and outline the process of generating the training dataset from MHD simulations. Section~\ref{Results} provides an evaluation of our diffusion model's performance in predicting magnetic field strength, comparing it to traditional DCF methods. 
Finally, we summarize our findings and conclusions in Section~\ref{Conclusions}.

\section{Data and Method}
\label{Data and Method}

\subsection{Magnetohydrodynamics Simulations}
\label{Magnetohydrodynamics Simulations}

\begin{figure*}[hbt!]
\centering
\includegraphics[width=0.75\linewidth]{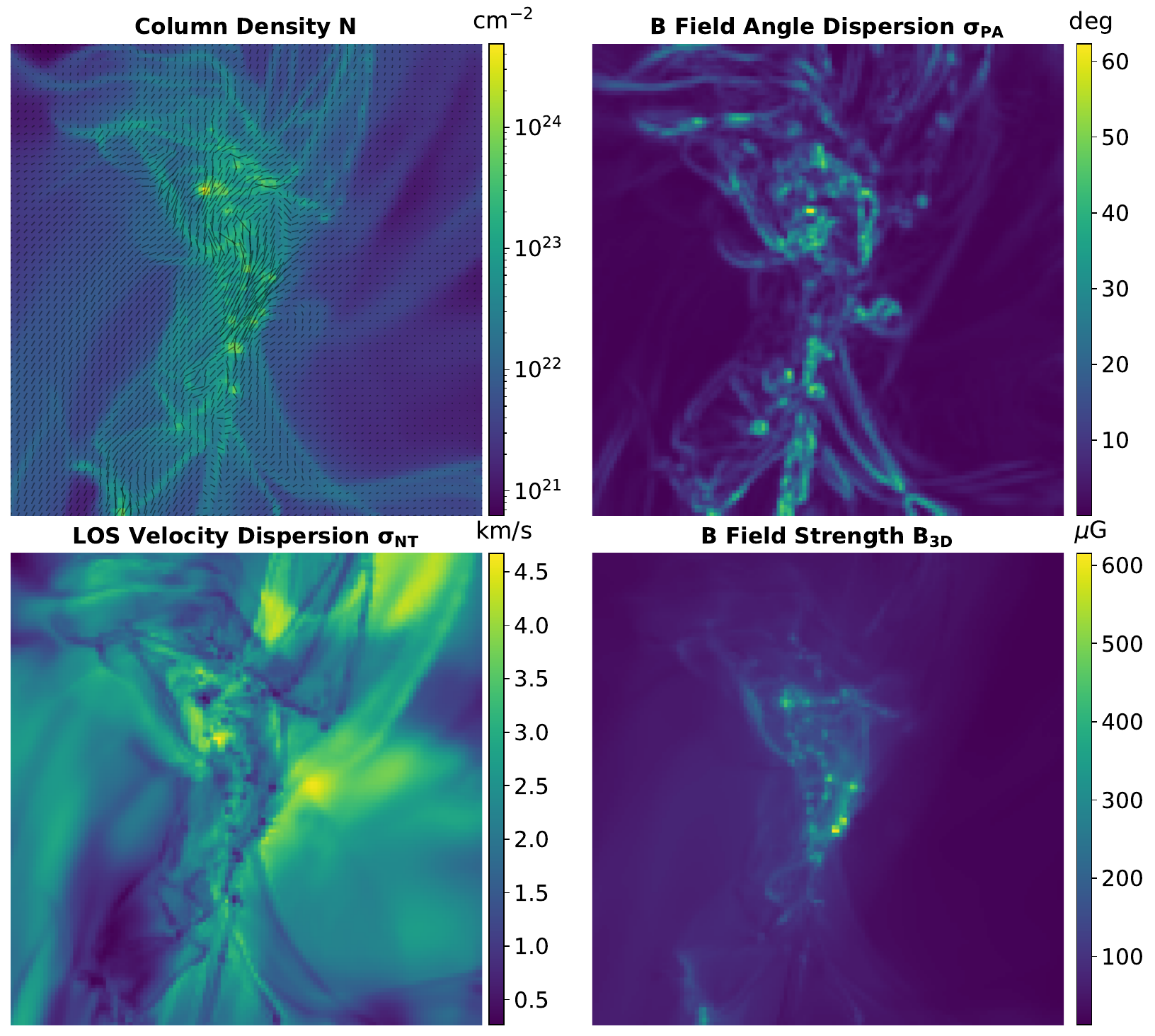}
\caption{Example of molecular cloud data for machine learning training: column density map with magnetic field directions (top left), magnetic field angle dispersion (top right), LOS velocity dispersion (bottom left), and projected magnetic field strength (bottom right).  }
\label{fig.simulation_all_N_angle_B}
\end{figure*} 

We conduct ideal MHD simulations based on the setups from \citet{2020ApJ...891..168W} and \citet{2023MNRAS.522..700H} using the MUSCL-Dedner method and HLLD Riemann solver within the adaptive mesh refinement (AMR) code Enzo \citep{2002JCoPh.175..645D, 2009ApJ...696...96W, 2014ApJS..211...19B}. These simulations include self-gravity, magnetic fields, and heating/cooling based on a photo-dissociation model, assuming a FUV radiation field of $G_0=4$ Habings, attenuation following the $n_\textrm{H}-A_V$ relation from \citet{2015ApJ...811...56W}, and a cosmic ray ionization rate of $\zeta=10^{-16}\,{\rm s^{-1}}$. The setup includes two clouds, each with a radius of 20~pc, initialized in a 128~pc$^3$ domain with a resolution of 256$^3$ cells. The clouds have an initial density of $n_\textrm{H}=83\:\textrm{cm}^{-3}$, a temperature of $T=15\:$K, and a solenoidal turbulent velocity field with $v_k^2 \propto k^{-4}, 2 \leq k \leq 20$. The surrounding gas has a density 10 times lower and a temperature 10 times higher to maintain pressure balance. While the GMCs begin with a temperature of 15~K, a multiphase temperature structure soon forms, with typical temperatures of $\sim 10-20\:$K at high densities ($n_{\rm H}\gtrsim 10^3\:{\rm cm}^{-3}$), $\sim 40\:$K at intermediate densities ($n_{\rm H}\sim 10^2\:{\rm cm}^{-3}$), and $\sim 1,000\:$K at low densities ($n_{\rm H}\lesssim 10\:{\rm cm}^{-3}$) \citep{2023MNRAS.522..700H}.

The initial magnetic field is set at a 60$^\circ$ angle to the collision axis, with strengths of 10, 30, and 50~$\mu$G across different cases. Four additional refinement levels are employed to resolve the local Jeans length with 8 cells. For each magnetic field strength, we model two GMC setups: non-colliding and colliding. In the colliding cases, the clouds have a relative velocity of 10 \kms and are offset by 0.5$\:R_{\rm GMC}$. These simulations, which run for 4.1 Myr, do not include star formation or feedback, thus representing the early phases of collapse before star formation begins. We analyze 22 evolutionary stages between 2 and 4.1~Myr, with 0.1~Myr intervals.

To enhance the diversity of the data set, we generate column density maps and their corresponding LOS mass-weighted polarization angle and their corresponding LOS nonthermal velocity dispersion and their corresponding projected true 3D magnetic field strength across different scales by adopting different AMR levels with different physical resolutions. {It is important to note that the LOS nonthermal velocity dispersion is influenced not only by turbulence but also by large-scale motions, e.g., cloud collisions, which contribute to the overall velocity dispersion. We do not exclude these effects, as it is challenging to disentangle them in real observations. Therefore, we replicate the conditions of observational data in our training set by including all contributions directly.} The image size in pixels is $128\times128$, with multiple physical scales, including 32, 16, 8, and 4 pc. In total, we have 25,479 images in the data set, in which 70\% are used for the training set, and the remaining 30\% are a test set. Additionally, these images are seen from a random direction of the viewing angle to increase the diversity of the data set. Figure~\ref{fig.simulation_all_N_angle_B} provides an example of the molecular cloud data used in our training set.

\subsection{Denoising Diffusion Probabilistic Models}
\label{Denoising Diffusion Probabilistic Models}

\begin{figure*}[hbt!]
\centering
\includegraphics[width=0.99\linewidth]{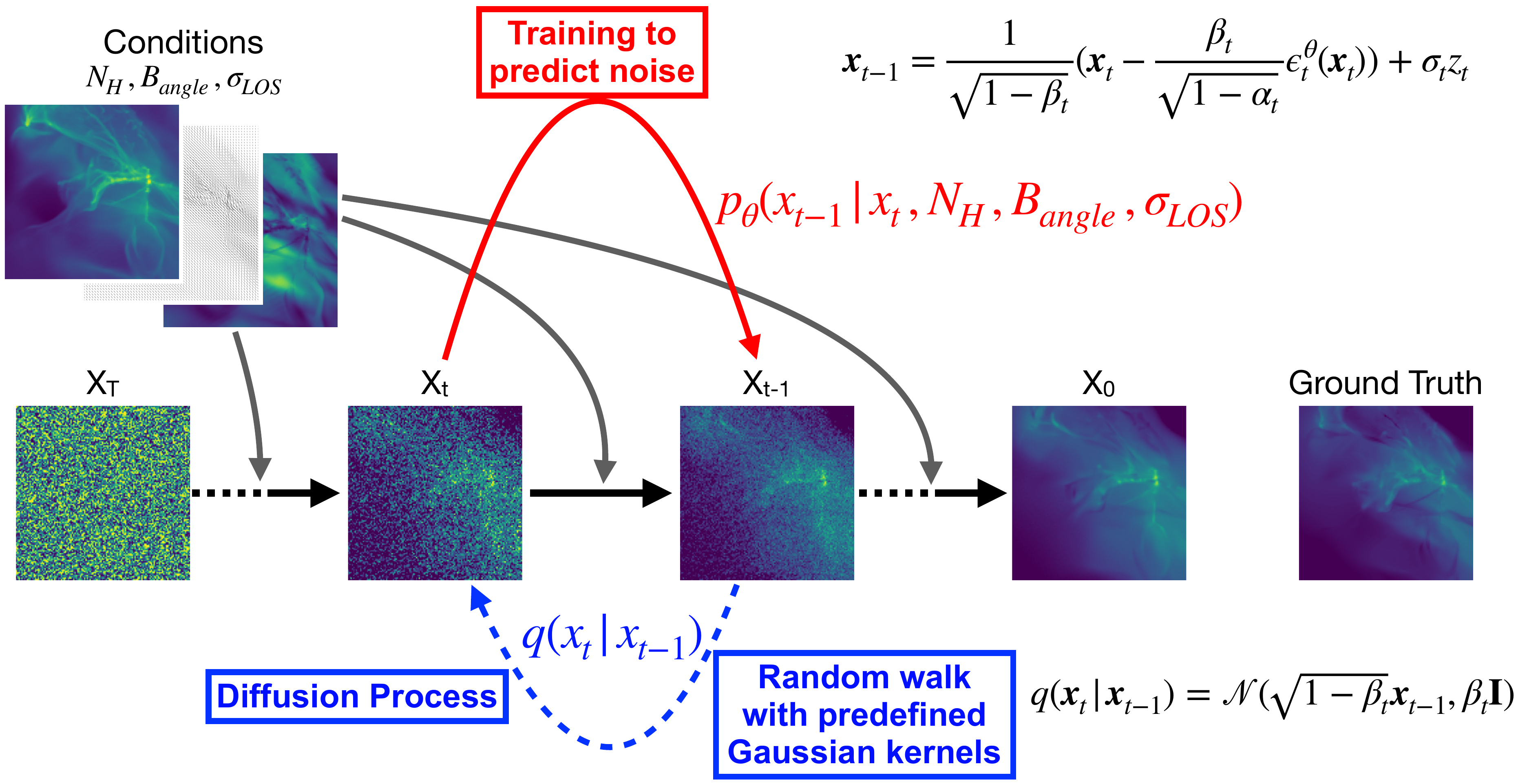}
\caption{Schematic workflow of the DDPM, illustrating both the diffusion and denoising processes. }
\label{fig.ddpm_B_plot}
\end{figure*} 

Denoising Diffusion Probabilistic Models (DDPMs) have emerged as a cutting-edge framework in deep learning, particularly in generative modeling, where they achieve state-of-the-art performance in capturing and reconstructing complex data distributions \citep{pmlr-v37-sohl-dickstein15,NEURIPS2020_diffusion,stablediff}. These models utilize principles from probability theory and stochastic processes to generate data by progressively modeling the statistical structure of the target distribution. Unlike traditional approaches, DDPMs excel at learning intricate, high-dimensional, and nonlinear relationships, making them particularly effective for domains where explicit analytical models are intractable.

At the core of DDPMs lies a forward-and-reverse diffusion mechanism. The forward process incrementally corrupts data by adding Gaussian noise through a sequence of steps, effectively transforming the data into a simple Gaussian distribution. Conversely, the reverse process, parameterized by a deep neural network (often a U-Net architecture), reconstructs the original data by iteratively removing the noise. Through this denoising sequence, DDPMs recover the true underlying structure of the data while minimizing unwanted perturbations. The entire process is governed by a predefined variance schedule, ensuring smooth transitions along the diffusion path. This framework enables DDPMs to model highly complex distributions with remarkable precision, as the progressive nature of the diffusion process allows for finer-grained reconstructions.

The goal of generative modeling in DDPMs can be mathematically described as learning a mapping between two distributions: a simple, easy-to-sample prior distribution $X\sim\mathcal{N}(0,I_d)$, and the target data distribution $Y$, which often represents a high-dimensional, intricate dataset such as natural images. In the context of machine learning, the dimensionality of $Y$ can be extraordinarily large. DDPMs learn this mapping by training a model $p_\theta$ to approximate samples from $q(y)$, the true data distribution. Once trained, the model can generate new samples by drawing from $\mathcal{N}(0,I_d)$ and transforming the noise through the learned reverse diffusion process to approximate the manifold of $Y$.

Conditional DDPMs extend this framework by incorporating auxiliary information, such as physical parameters or observational constraints, to guide the generative process. By conditioning the model on additional inputs, DDPMs can integrate domain-specific knowledge, enabling them to produce outputs that are both realistic and consistent with physical laws. This conditional design is particularly useful in applications such as astrophysical modeling, where the relationships between variables are often highly non-linear and difficult to parameterize analytically. For example, in astrophysics, DDPMs can condition on observational data like column density, polarization angles, or line-of-sight velocity dispersion to predict physical quantities such as magnetic field strength. By combining domain-specific priors with the model’s ability to handle high-dimensional, non-linear mappings, DDPMs provide a robust framework for simulating, analyzing, and interpreting complex astrophysical phenomena.

The adaptability and precision of DDPMs make them particularly well-suited for astrophysical research. In many cases, astrophysical systems involve high-dimensional data distributions and subtle dependencies between physical variables that are not easily captured by traditional modeling techniques. By leveraging DDPMs, researchers can explore these complex relationships with unprecedented accuracy, reducing the need for oversimplifications and approximations. Moreover, DDPMs can be trained on synthetic data generated from simulations, allowing them to generalize to real-world observations while accounting for diverse and intricate physical processes. Their ability to seamlessly integrate domain-specific conditions further enhances their applicability in bridging theoretical models and observational data in fields like astronomy and astrophysics.

The schematic workflow of DDPM is illustrated in Figure~\ref{fig.ddpm_B_plot}. Unlike traditional discriminative models, which focus on defining decision boundaries (e.g., CNNs for recognition tasks), DDPMs are generative models that aim to learn the complete data distribution. This capability enables them to generate new samples by mimicking the original distribution rather than merely distinguishing classes.

The loss function used in training DDPMs has seen several advancements in recent research. In this work, we employ the original DDPM formulation introduced by \citet{NEURIPS2020_diffusion}, which utilizes a variational lower bound (VLB) loss--a common objective in probabilistic generative models like variational autoencoders (VAEs). The VLB loss function, defined as follows, maximizes the likelihood of the data by optimizing a lower bound on its probability, thus guiding the model toward a smooth and stable generative process:
\begin{align}
&\mathbb{E}_{q}[D_{KL}(q(\mathbf{x}_{T}|\mathbf{x_0})||p_{\theta}(x_T)) + \nonumber \\
&\sum_{t>1}D_{KL}(q(\mathbf{x}_{t-1}|\mathbf{x}_{t},\mathbf{x}_{0})||p_{\theta}(\mathbf{x}_{t-1}|\mathbf{x}_{t})) - \text{log}p_{\theta}(\mathbf{x}_0|\mathbf{x}_1)],
\end{align}
where $q$ represents the forward diffusion process parameterized by a sequence of Gaussian noises, and $p_\theta$ denotes the neural network with learnable parameters $\theta$. In practice, however, the model is implemented using a U-Net backbone, and training is simplified by minimizing the mean squared error (MSE) between the predicted noise and the noise defined by the scheduler. This MSE-based objective aligns with the diffusion process, making training more straightforward by directly matching the predicted and target noise distributions, ultimately enhancing the model’s stability and performance.

In this work, we adopt the same diffusion model outlined in \citet{2023ApJ...950..146X}, which provides a detailed mathematical explanation of the DDPM formulation. We consider three different tasks:
\begin{enumerate}
\item The first task uses a single input channel (condition) for the DDPM to infer the magnetic field strength, based solely on column density.
\item The second task involves two input channels, incorporating both column density and polarization angle.
\item The third task utilizes all three channels of information, including column density, polarization angle, and LOS nonthermal velocity dispersion.
\end{enumerate}

These tasks are designed to accommodate different observational scenarios, where certain data may be unavailable, such as missing polarization angle measurements or molecular line data for LOS velocity dispersion. For each task, we train three different models and evaluate their performance:
\begin{enumerate}
\item One model is trained on all initial conditions, covering all magnetic field strengths and both colliding and non-colliding GMC scenarios.
\item Another model is trained only on colliding GMC scenarios and tested on non-colliding ones.
\item An additional model is trained only on the 10~$\mu$G and 50~$\mu$G cases and tested on the 30~$\mu$G case.
\end{enumerate}

In total, we have 25,479 data samples, with 75\% allocated for training and the remaining 25\% for testing. Our DDPM training was conducted on a single NVIDIA TITAN V GPU, which, while not the most powerful by current standards, is a cost-efficient consumer-grade option. We set the training to run for 600 epochs, and due to the intrinsic configuration of generative models like DDPM, the optimal batch size was 1, resulting in a total of 11,465,400 iterations over the data. The total training time for each model was 248 hours, or approximately 10 days. The performance of each model is presented in Section~\ref{Evaluation of DDPM Performance}.


\section{Results}
\label{Results}

\subsection{$B-N$ Relation}
\label{sec.B-N relation}

\begin{figure*}[hbt!]
\centering
\includegraphics[width=0.99\linewidth]{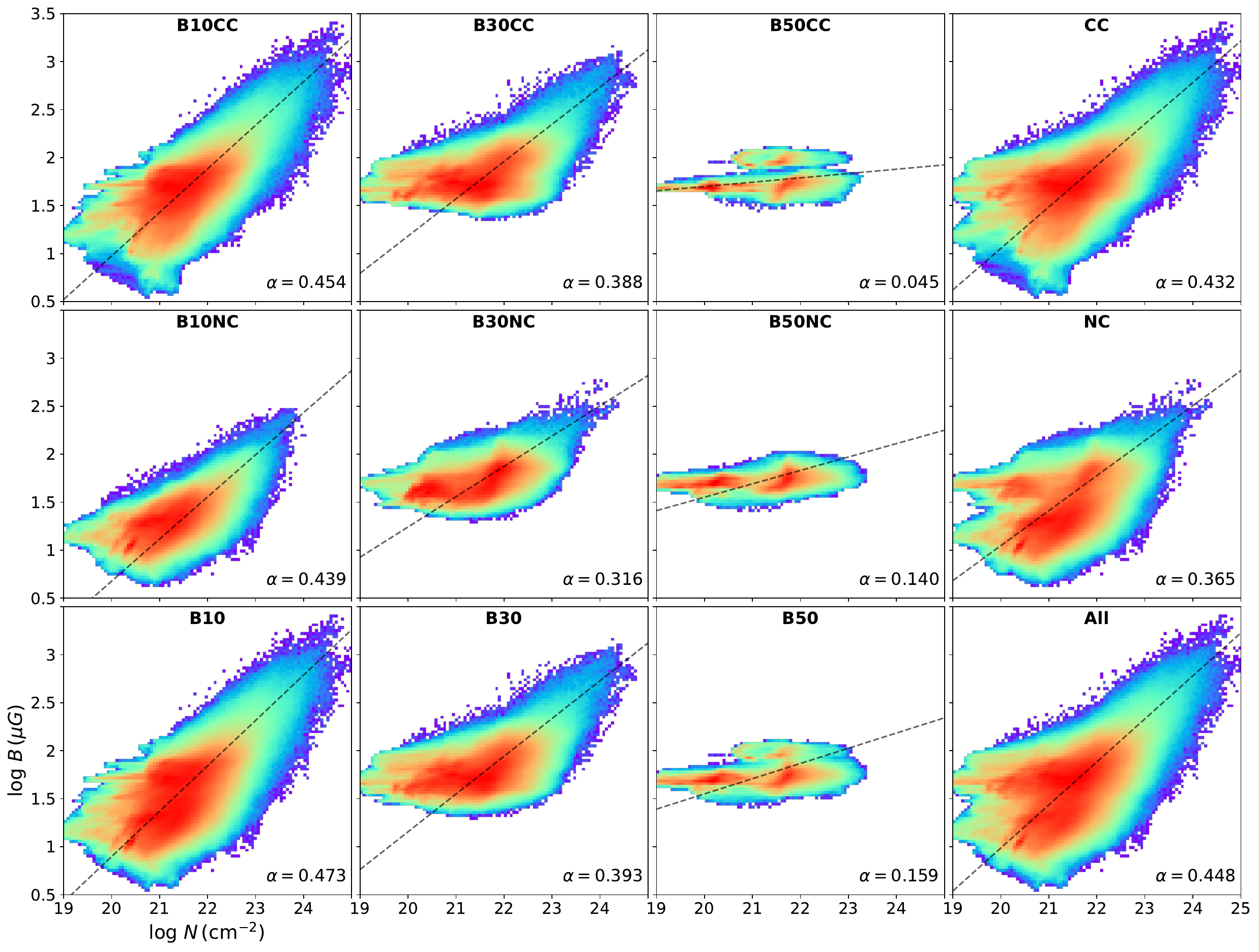}
\caption{Correlation between magnetic field strength and column density under different physical conditions. From the first to third columns: initial magnetic field strengths of 10 $\mu$G, 30 $\mu$G, and 50 $\mu$G, respectively. The first and second rows represent the colliding and non-colliding cloud scenarios. The third row combines both colliding and non-colliding cloud scenarios for the three different initial magnetic field strengths. The fourth column (1st and 2nd rows) shows the combined results for all three initial magnetic field strengths in the colliding and non-colliding scenarios, respectively. The bottom-right panel displays the correlation between magnetic field strength and column density across all simulation data. The dashed line represents the best-fit power-law for each panel, with the power-law exponent shown in the bottom-right corner of each panel.}
\label{fig.B_N_hist_img_all}
\end{figure*} 


\begin{figure}[hbt!]
\centering
\includegraphics[width=0.99\linewidth]{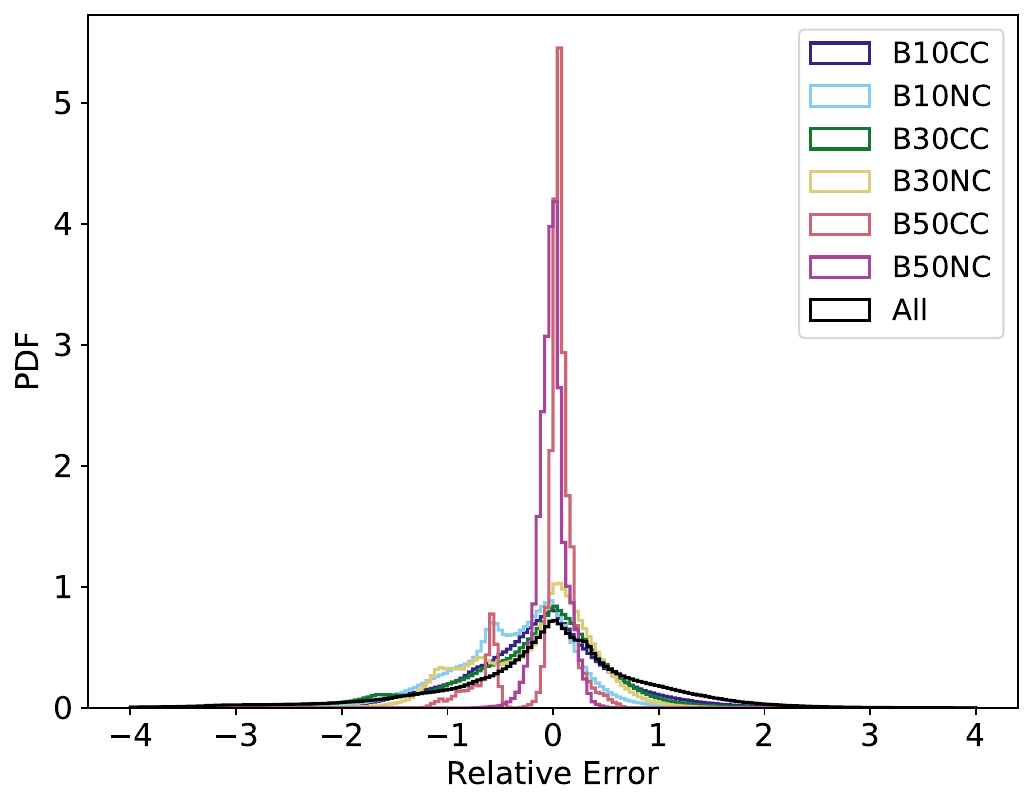}
\caption{Probability distribution function of the relative error ($\delta_{B}$) in the inferred magnetic field strength from the column density, based on the best-fit power-law for each specific physical condition.}
\label{fig.B_powerlaw_fit_error_all}
\end{figure} 

The relationship between magnetic field strength ($B$) and number density ($n$), commonly referred to as the $B-n$ relation, is a central focus in both observational studies \citep{1993ApJ...407..175C,1999ApJ...520..706C,2005ApJ...624..773H,2010ApJ...725..466C,2011ApJ...728..146Y,2021ApJ...917...35M} and numerical simulations \citep{2001ApJ...546..980O,2015MNRAS.452.2500L,2016ApJ...831...85Y,2023ApJ...946L..46C}. \citet{1966MNRAS.133..265M} first proposed a theoretical framework linking $B$ and $n$ through studies of collapsing, magnetized, gravitationally bound clouds, suggesting power-law relations: $B \propto n^{2/3}$ for weak fields and $B \propto n^{1/2}$ for strong fields. Observationally, \citet{2010ApJ...725..466C} employed Zeeman surveys of HI, OH, and CN spectral lines to measure the LOS magnetic field strength ($B_z$), finding $B_z \propto n^{0.65}$. Likewise, \citet{2021ApJ...917...35M} analyzed 17 dense cores using the DCF method to estimate the POS magnetic field strength ($B_{POS}$), reporting an exponent of 0.66. However, estimating the number density in molecular clouds is challenging in observations. Therefore, we examine the correlation between magnetic field strength and column density ($N$), a more readily obtained observable. 

Figure~\ref{fig.B_N_hist_img_all} illustrates the relationship between magnetic field strength and column density under different initial magnetic field strengths and dynamic conditions. The best-fit power-law for each scenario is also shown. While there is significant scatter in the $B-N$ relations, correlation exist, suggesting that magnetic field strength can potentially be inferred from column density. The figure also provides the power-law exponents, showing that these vary depending on physical conditions. When averaged across all conditions, the power-law exponent is 0.448.


In Figure~\ref{fig.B_powerlaw_fit_error_all}, we illustrate the relative error in estimating magnetic field strength based on the corresponding power-law fit for each scenario, using column density. The relative error ($\delta_{B}$) is defined symmetrically as follows:
\begin{equation} \label{eqn_rela_error}
\delta_{B}=\frac{B_{Pred}-B_{True}}{min(B_{Pred},B_{True})}.
\end{equation}
This formulation provides a more balanced approach compared to the classical definition of relative error, which uses $B_{True}$ as the denominator. In the classical method, when the predicted value overestimates, it is easy to assess the factor of overestimation. However, in cases of underestimation, the error falls between -1 and 0, making it difficult to gauge how much the predicted value differs from the true one. By defining $\delta_{B}$ symmetrically, we can more clearly interpret the factor difference between predicted and true values. For instance, if the prediction underestimates by 50\%, $\delta_{B}$ returns a value of -1, indicating that the true value is a factor of 1 larger than the predicted value. Simulations with an initial magnetic field strength of 50 $\mu$G show the smallest dispersion, likely due to the narrower dynamic range of magnetic field strength in these cases, where strong magnetic support prevents significant gravitational collapse.

\subsection{Evaluation of DDPM Performance}
\label{Evaluation of DDPM Performance}

\begin{figure*}[hbt!]
\centering
\includegraphics[width=0.99\linewidth]{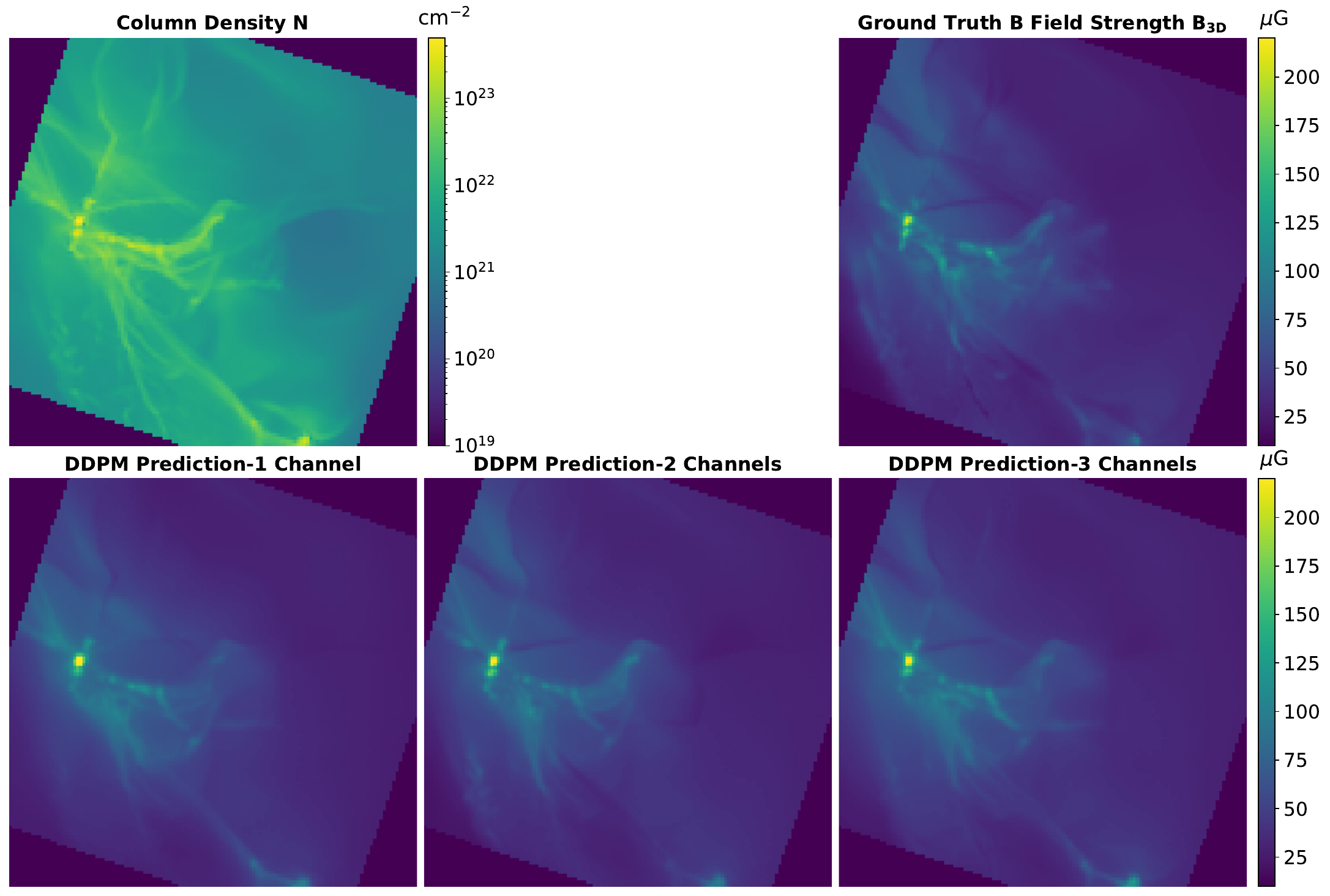}
\caption{An example showcasing the performance of three models trained with different inputs: single-channel (column density), two-channel (column density + polarization angle), and three-channel (column density + polarization angle + LOS nonthermal velocity dispersion) in predicting the magnetic field strength of a piece of molecular cloud.}
\label{fig.ddpm_pred_img_channel_test_all}
\end{figure*} 

\begin{figure*}[hbt!]
\centering
\includegraphics[width=0.99\linewidth]{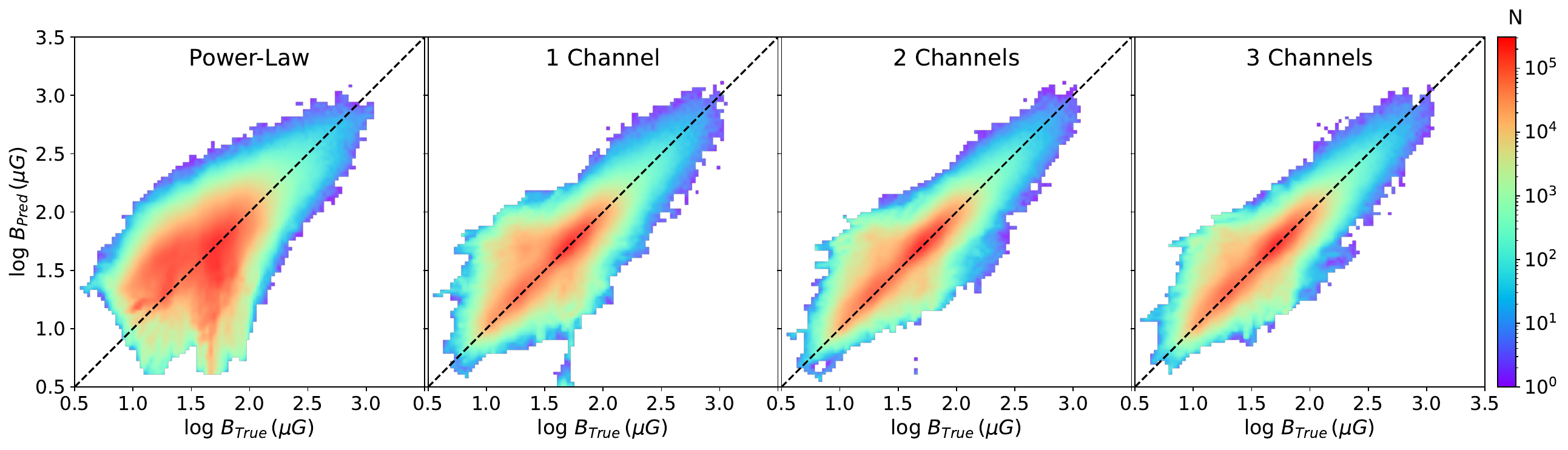}
\caption{2D histograms comparing the ground truth magnetic field strength with inferred values across all physical condition samples: using power-law fitting from column density (1st panel), and from the three different trained DDPMs (2nd-4th panels). }
\label{fig.ddpm_pred_hist_channel_test_all}
\end{figure*} 

\begin{figure}[hbt!]
\centering
\includegraphics[width=0.99\linewidth]{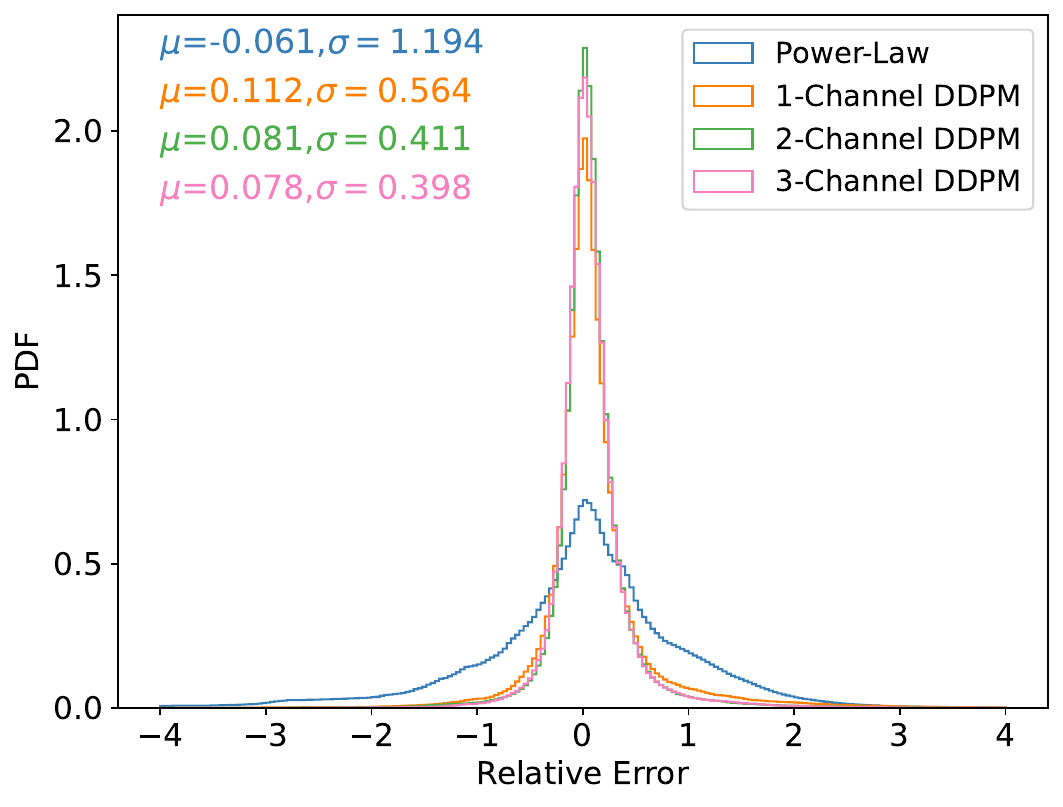}
\caption{Relative error ($\delta_{B}$) distribution between predicted and true values across all physical condition samples for different models. }
\label{fig.ddpm_pred_error_plot_test_all}
\end{figure} 

\begin{figure*}[hbt!]
\centering
\includegraphics[width=0.99\linewidth]{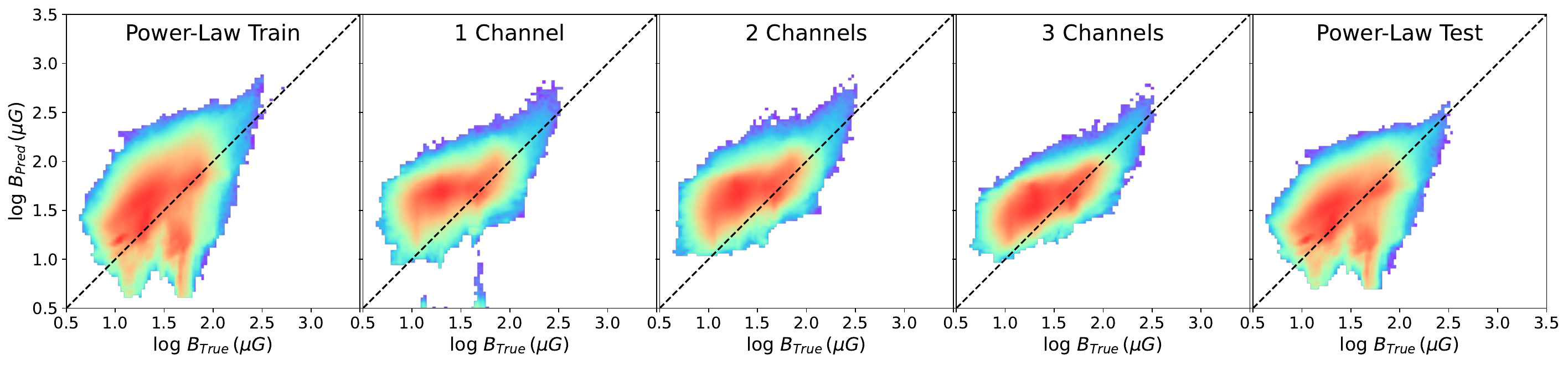}
\caption{2D histograms comparing the ground truth magnetic field strength with the inferred values for the test set (non-colliding GMC scenarios): using power-law fitting based on column density from the training set (colliding GMC scenarios) in the 1st panel, and from three different trained DDPMs trained on colliding GMC scenarios in the 2nd-4th panels. The 5th panel shows the power-law fitting based on column density from the test set (non-colliding GMC scenarios). }
\label{fig.ddpm_pred_hist_channel_test_NC}
\end{figure*} 

\begin{figure}[hbt!]
\centering
\includegraphics[width=0.99\linewidth]{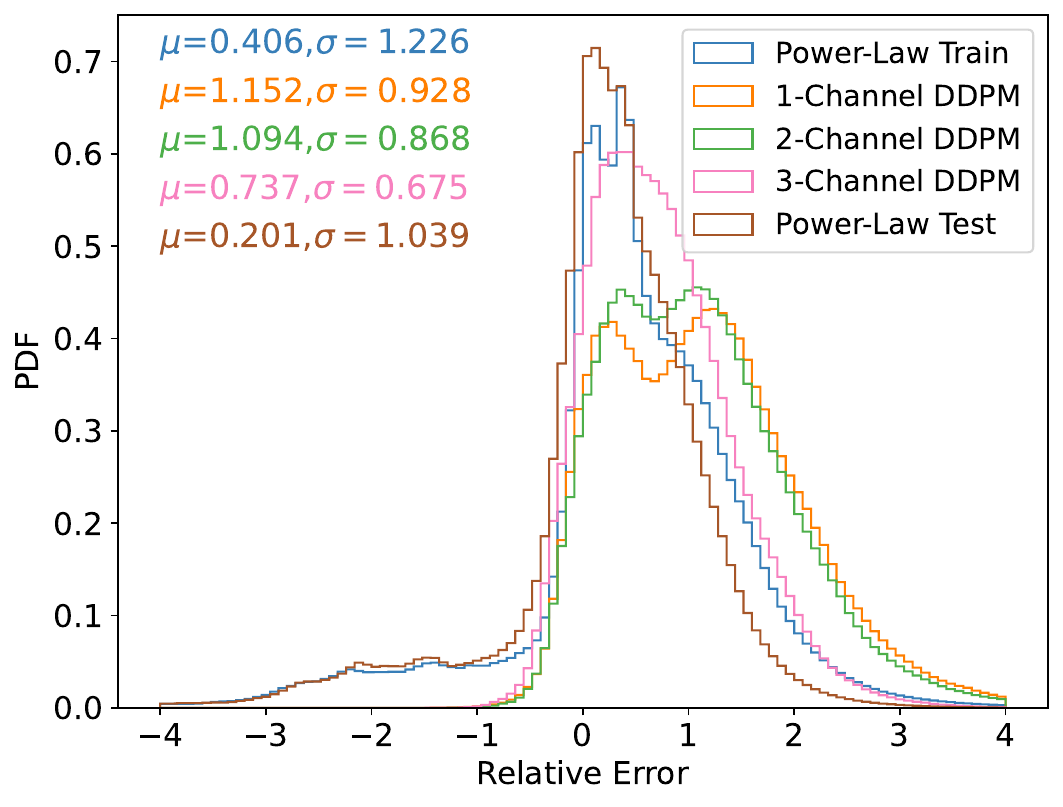}
\caption{Relative error ($\delta_{B}$) distribution between the predicted and true values for the non-colliding test samples across different models. }
\label{fig.ddpm_pred_error_plot_test_NC}
\end{figure} 

\begin{figure*}[hbt!]
\centering
\includegraphics[width=0.99\linewidth]{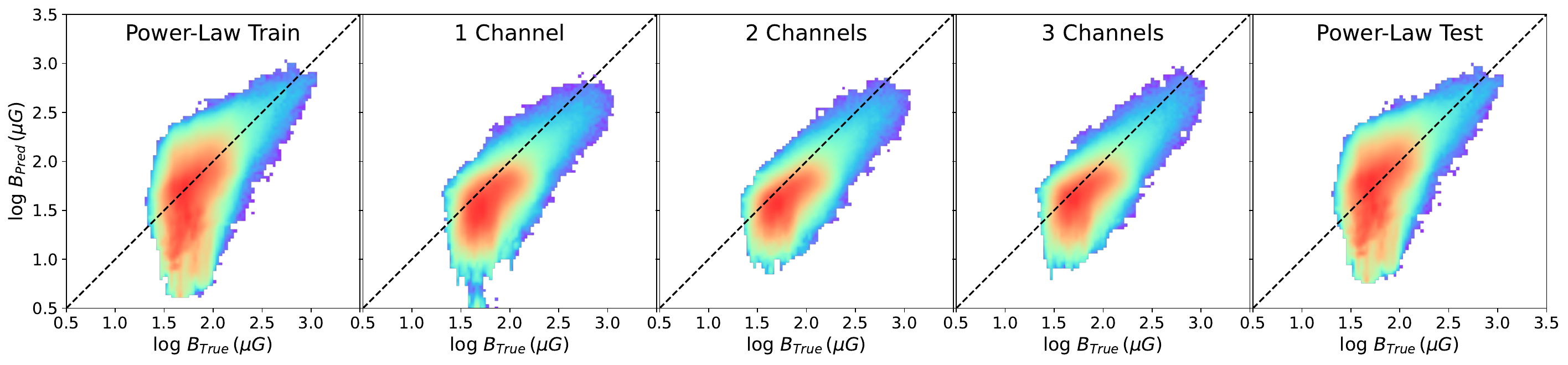}
\caption{2D histograms comparing the ground truth magnetic field strength with the inferred values for the test set (30~$\mu$G): using power-law fitting based on column density from the training set (10~$\mu$G and 50~$\mu$G cases) in the 1st panel, and from three different trained DDPMs trained on colliding GMC scenarios in the 2nd-4th panels. The 5th panel shows the power-law fitting based on column density from the test set (30~$\mu$G). }
\label{fig.ddpm_pred_hist_channel_test_B30}
\end{figure*} 

\begin{figure}[hbt!]
\centering
\includegraphics[width=0.99\linewidth]{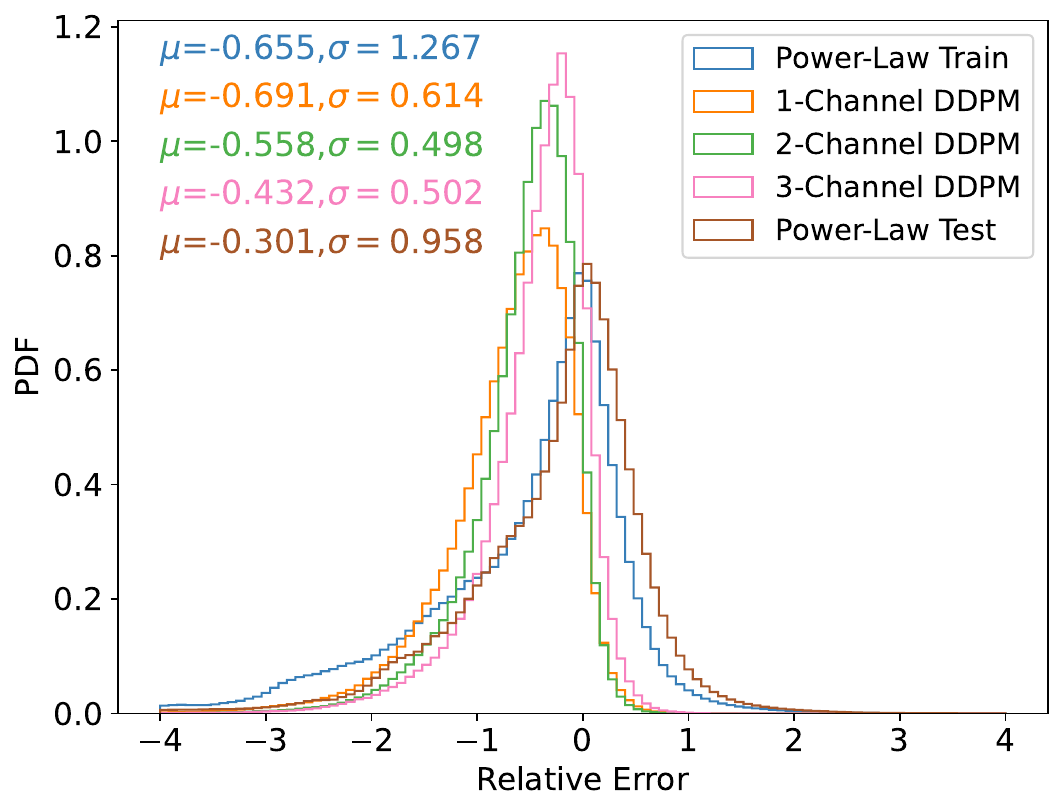}
\caption{Relative error ($\delta_{B}$) distribution between the predicted and true values for the 30~$\mu$G case test samples across different models. }
\label{fig.ddpm_pred_error_plot_test_B30}
\end{figure} 

In this section, we evaluate the performance of the trained DDPMs on the test sets. We first examine the models trained on all initial conditions, which encompass a range of magnetic field strengths and both colliding and non-colliding GMC scenarios. Figure~\ref{fig.ddpm_pred_img_channel_test_all} illustrates an example of predictions made by three models trained with different inputs: single-channel (column density), two-channel (column density + polarization angle), and three-channel (column density + polarization angle + LOS nonthermal velocity dispersion). To quantitatively assess these results, Figure~\ref{fig.ddpm_pred_hist_channel_test_all} presents 2D histograms comparing the ground truth magnetic field strength to the DDPM-predicted values for each model, with power-law fitting inferred from column density also shown for comparison. To further evaluate uncertainty, Figure~\ref{fig.ddpm_pred_error_plot_test_all} displays the relative error ($\delta_{B}$, Equation~\ref{eqn_rela_error}) distributions between predicted and true values for all tested samples, and provides the mean and standard deviation of these relative errors.


Next, we evaluate the models trained on a subset of physical conditions and tested on new, unseen conditions. Figure~\ref{fig.ddpm_pred_hist_channel_test_NC} shows 2D histograms comparing the ground truth magnetic field strength with the inferred values for the test set (non-colliding GMC scenarios), including power-law fitting based on column density from both the training set (colliding GMC scenarios) and the test set (non-colliding GMC scenarios). The relative error ($\delta_{B}$) distributions for these non-colliding test samples across different models are shown in Figure~\ref{fig.ddpm_pred_error_plot_test_NC}. Notably, applying the power-law fitting from the training set (colliding GMC scenarios) to the test set (non-colliding scenarios) leads to a clear overestimation of magnetic field strength. A similar trend is observed in the DDPMs trained on colliding scenarios. This discrepancy, highlighted by the distinct difference in exponents from the $B-N$ relation between colliding and non-colliding GMC scenarios (Figure~\ref{fig.B_N_hist_img_all}), suggests that while DDPMs are powerful, they may not generalize perfectly to physical conditions not represented in the training data, warranting caution when applying them to new scenarios.

Finally, we evaluate the models trained on the 10~$\mu$G and 50~$\mu$G cases and tested on the 30~$\mu$G case. Figure~\ref{fig.ddpm_pred_hist_channel_test_B30} presents 2D histograms comparing the ground truth magnetic field strength to the inferred values for the test set (30~$\mu$G), including power-law fitting based on column density from both the training set (10~$\mu$G and 50~$\mu$G cases) and the test set (30~$\mu$G). The relative error ($\delta_{B}$) distributions for these test samples are shown in Figure~\ref{fig.ddpm_pred_error_plot_test_B30}. Here, the DDPM trained with three channels performs significantly better than other methods, although some underestimation remains at the lower end of the magnetic field strength. Power-law fitting methods exhibit a long tail in relative error ($\delta_{B}$), whereas the DDPMs—regardless of the number of input channels—generally show less dispersion. This suggests that DDPMs are more capable of accurately inferring magnetic field strength for ``interpolated'' physical conditions, as the 30~$\mu$G case lies between the 10~$\mu$G and 50~$\mu$G training conditions. 

\subsection{Testing on New Simulations}
\label{Testing on New Simulations}

\begin{table*}[!htb]
\begin{center}
\caption{Summary of the Relative Error ($\delta_{B}$) for Different Models on New Simulations$^a$ \label{tab-model-orion-ddpm}}
    \begin{tabular}{cccccccccccccc}
    \hline
    \hline
        $\alpha_{\rm vir}$ & $\mu_{\Phi}$ & $B$ & $M_{A}$ & \multicolumn{2}{c}{PL-Train} & \multicolumn{2}{c}{1-Channel} & \multicolumn{2}{c}{2-Channel} & \multicolumn{2}{c}{3-Channel} &  \multicolumn{2}{c}{PL-Test}   \\ 
          &   &  ($\mu$G) &  & mean & std & mean & std & mean & std & mean & std & mean & std   \\ \hline 
        \multirow{5}{*}{2} & 1 & 25.5 & 0.87 & 0.339 & 0.324 & 0.788 & 0.675 & 0.294 & 0.322 & 0.213 & 0.505 & -0.095 & 0.186  \\ 
        & 2 & 12.8 & 1.75 & 0.673 & 0.526 & 1.364 & 1.040 & 0.134 & 0.545 & -0.147 & 0.286 & -0.207 & 0.446  \\ 
         & 4 & 6.4 & 3.50 & 1.209 & 0.778 & 2.073 & 1.226 & 0.556 & 0.762 & 0.105 & 0.405 & -0.105 & 0.470  \\ 
         & 8 & 3.2 & 6.99 & 1.450 & 0.855 & 2.507 & 1.723 & 0.573 & 0.804 & 0.196 & 0.442 & -0.027 & 0.534  \\ 
         & 16 & 1.6 & 13.98 & 2.287 & 1.143 & 3.672 & 2.281 & 1.301 & 1.200 & 0.644 & 0.647 & 0.000 & 0.646  \\ \hline 
         \multirow{5}{*}{1} & 1 & 51.1 & 0.62 & -0.259 & 0.623 & 0.115 & 0.330 & -0.076 & 0.202 & -0.076 & 0.301 & -0.238 & 0.469  \\ 
         & 2 & 25.5 & 1.24 & 0.285 & 0.478 & 0.552 & 0.537 & 0.160 & 0.394 & -0.184 & 0.396 & -0.149 & 0.417  \\ 
         & 4 & 12.8 & 2.47 & 1.331 & 0.732 & 1.735 & 0.913 & 1.149 & 0.643 & 0.340 & 0.561 & -0.227 & 0.526  \\ 
         & 8 & 6.4 & 4.94 & 1.194 & 0.783 & 1.587 & 1.001 & 0.819 & 0.721 & 0.213 & 0.579 & -0.169 & 0.545  \\ 
         & 16 & 3.2 & 9.89 & 1.765 & 0.948 & 2.675 & 1.635 & 1.338 & 1.092 & 0.720 & 0.867 & 0.026 & 0.551  \\ \hline    
\multicolumn{14}{p{0.5\linewidth}}{Notes:}\\
\multicolumn{14}{p{0.83\linewidth}}{$^a$ Virial parameter, mass-to-flux ratios, Alfv\'en mach number, and the mean and standard deviation of the relative error ($\delta_{B}$) for predictions from various models. From left to right: power-law fitting from the training set, 1-channel DDPM, 2-channel DDPM, 3-channel DDPM, and power-law fitting based on the column density from the corresponding new simulations (test set).}
\end{tabular}
\end{center}
\end{table*}

\begin{figure*}[hbt!]
\centering
\includegraphics[width=0.99\linewidth]{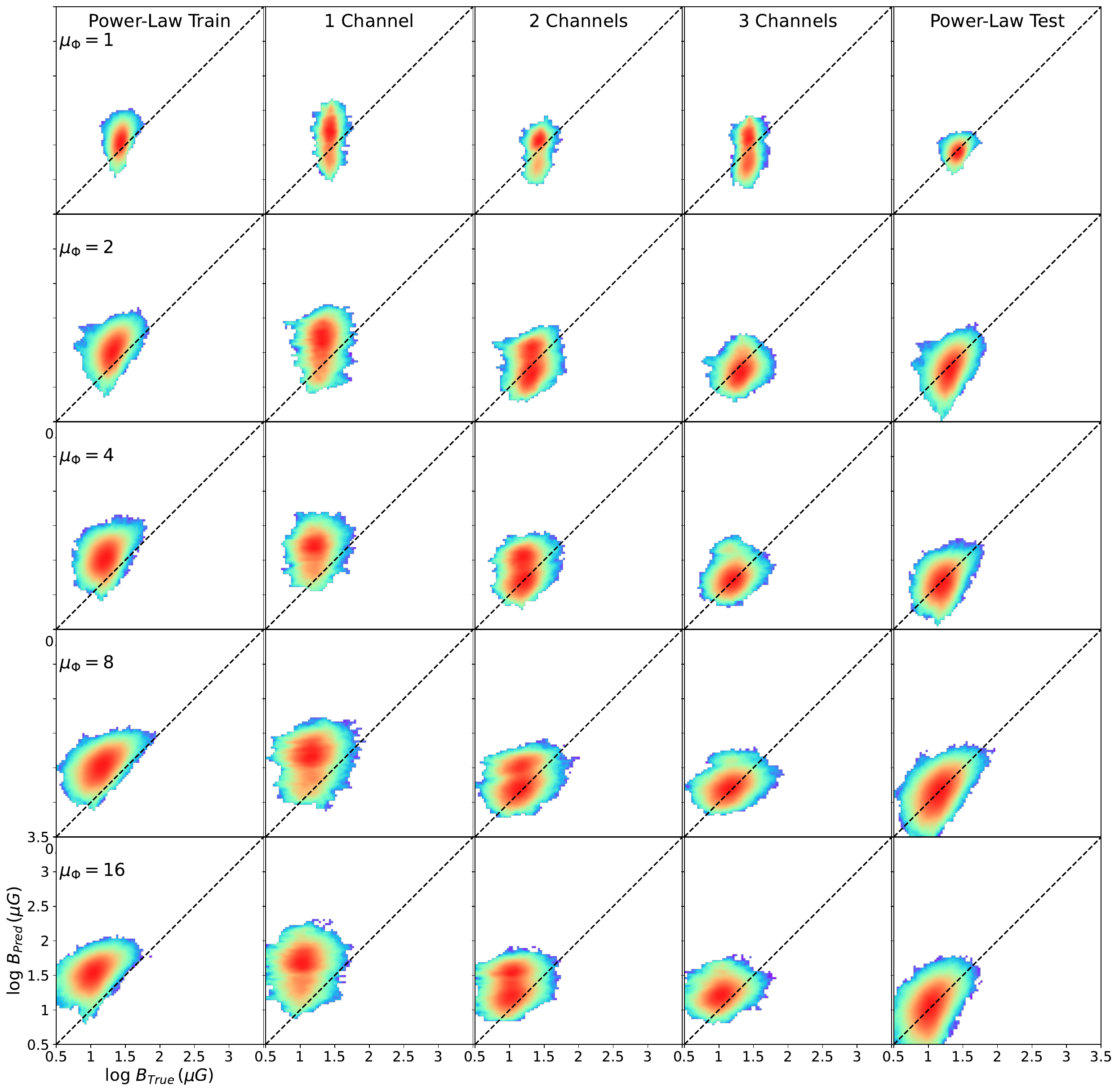}
\caption{2D histograms comparing the ground truth magnetic field strength with the inferred values for new simulations with a virial parameter $\alpha_{\rm vir}=2$ at different mass-to-flux ratios $\mu_{\Phi}$. From left to right: power-law fitting based on the training set (1st column), and predictions from three different trained DDPMs (2nd-4th columns). The 5th column presents power-law fitting based on the column density from the corresponding new simulations (test set).}
\label{fig.ddpm_pred_hist_channel_Orion_alpha2}
\end{figure*} 

\begin{figure*}[hbt!]
\centering
\includegraphics[width=0.99\linewidth]{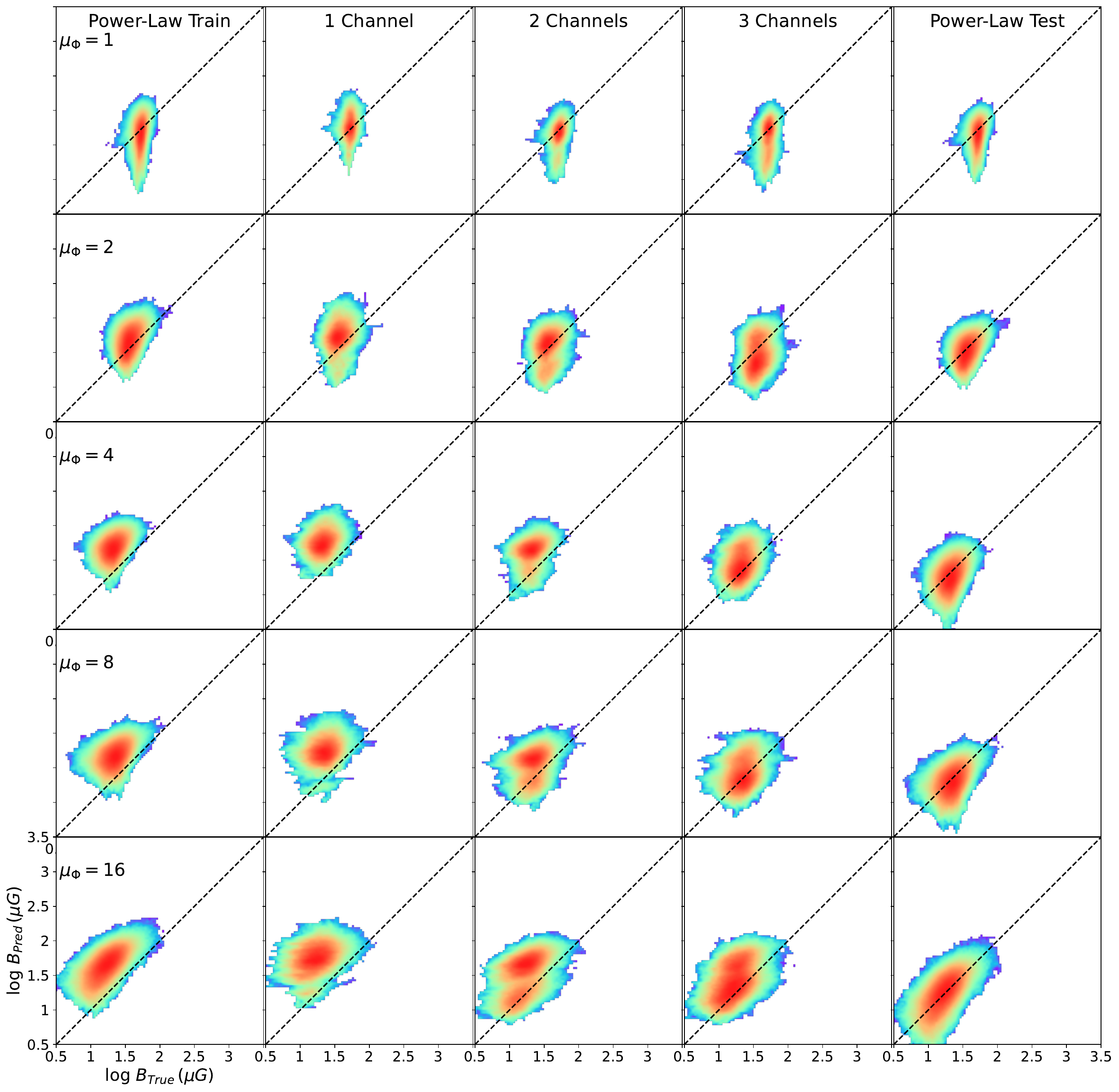}
\caption{Similar to Figure~\ref{fig.ddpm_pred_hist_channel_Orion_alpha2}, but for new simulations with a virial parameter of $\alpha_{\rm vir}=1$.}
\label{fig.ddpm_pred_hist_channel_Orion_alpha1}
\end{figure*} 

\begin{figure*}[hbt!]
\centering
\includegraphics[width=0.99\linewidth]{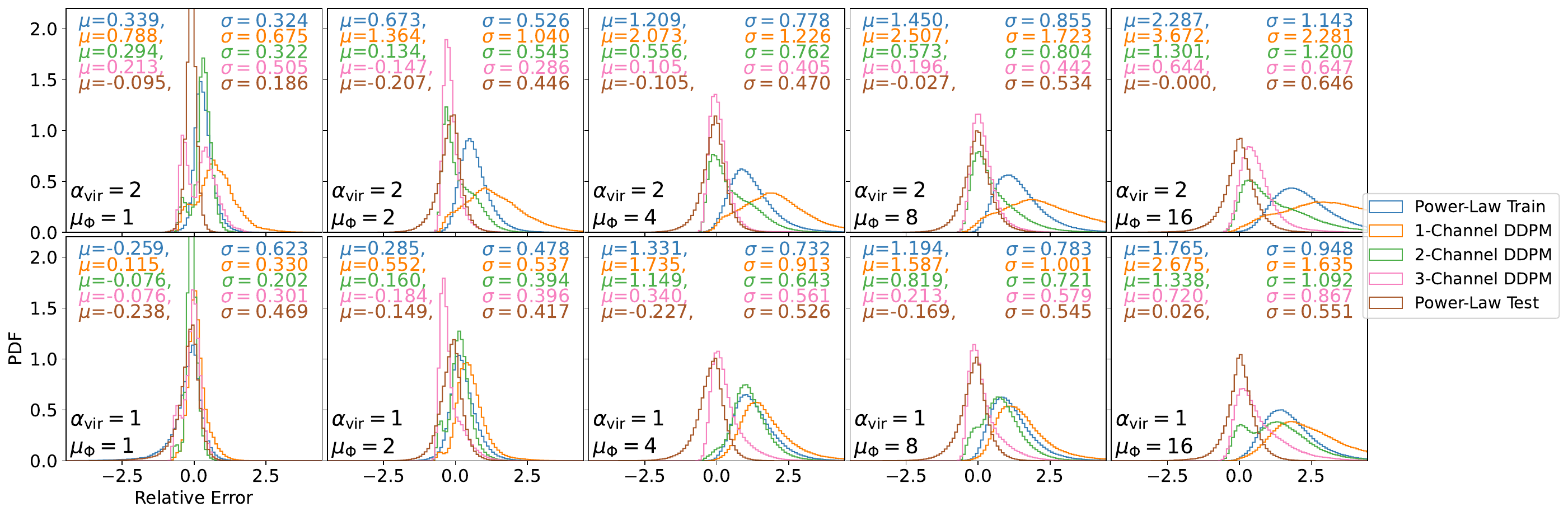}
\caption{Relative error ($\delta_{B}$) distribution between predicted and true values for the new simulations with different virial parameters ($\alpha_{\rm vir}$) and mass-to-flux ratios ($\mu_{\Phi}$). }
\label{fig.ddpm_pred_error_plot_test_Orion}
\end{figure*}

\begin{figure*}[hbt!]
\centering
\includegraphics[width=0.99\linewidth]{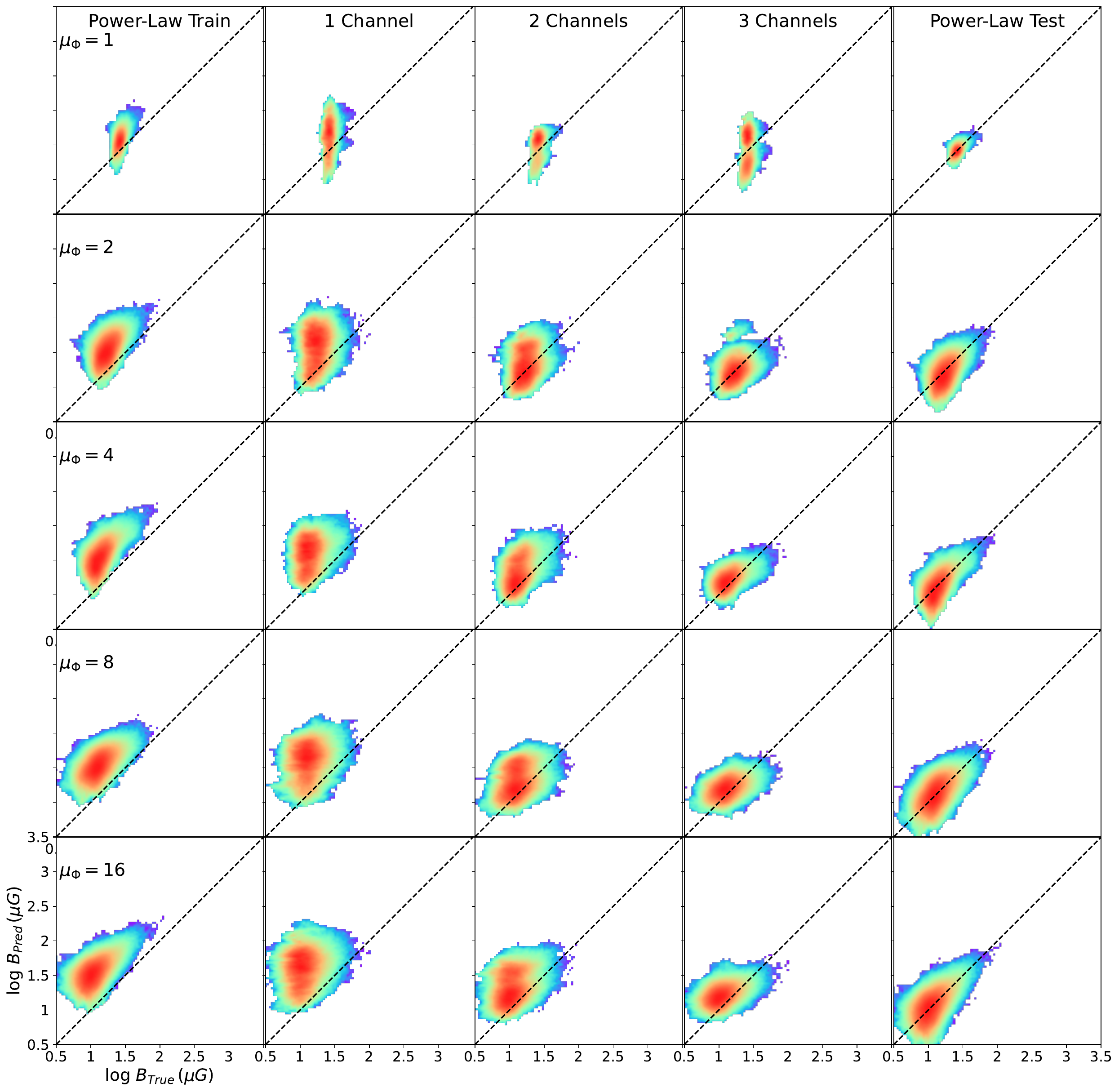}
\caption{Similar to Figure~\ref{fig.ddpm_pred_hist_channel_Orion_outflow_2p5_alpha2}, but showing results for new simulations that include self-gravity and outflow feedback, with a virial parameter of $\alpha_{\rm vir}=2$.}
\label{fig.ddpm_pred_hist_channel_Orion_outflow_2p5_alpha2}
\end{figure*} 

\begin{figure*}[hbt!]
\centering
\includegraphics[width=0.99\linewidth]{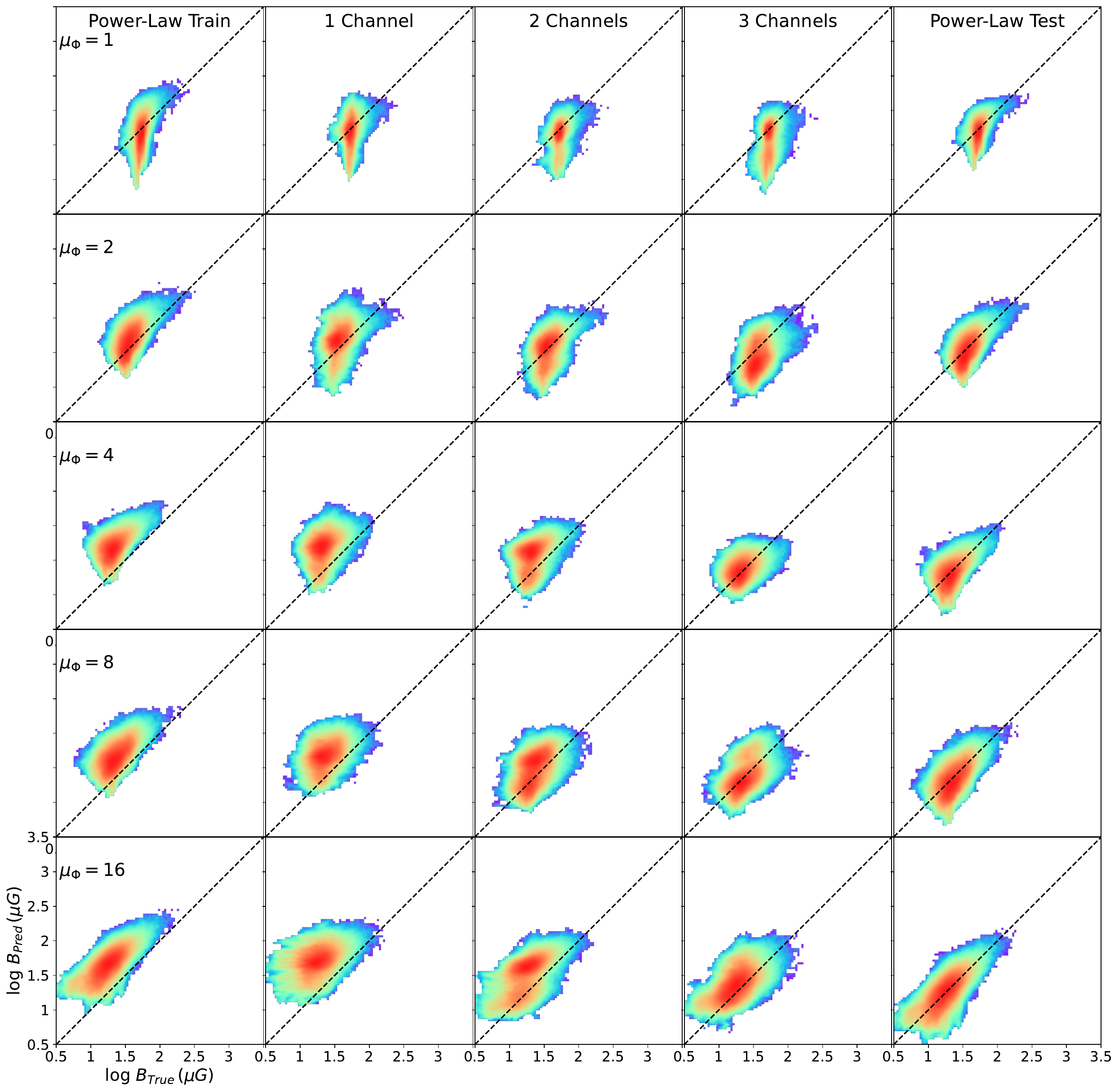}
\caption{Similar to Figure~\ref{fig.ddpm_pred_hist_channel_Orion_outflow_2p5_alpha2}, but showing results for new simulations that include self-gravity and outflow feedback, with a virial parameter of $\alpha_{\rm vir}=1$.}
\label{fig.ddpm_pred_hist_channel_Orion_outflow_2p5_alpha1}
\end{figure*} 

\begin{figure*}[hbt!]
\centering
\includegraphics[width=0.99\linewidth]{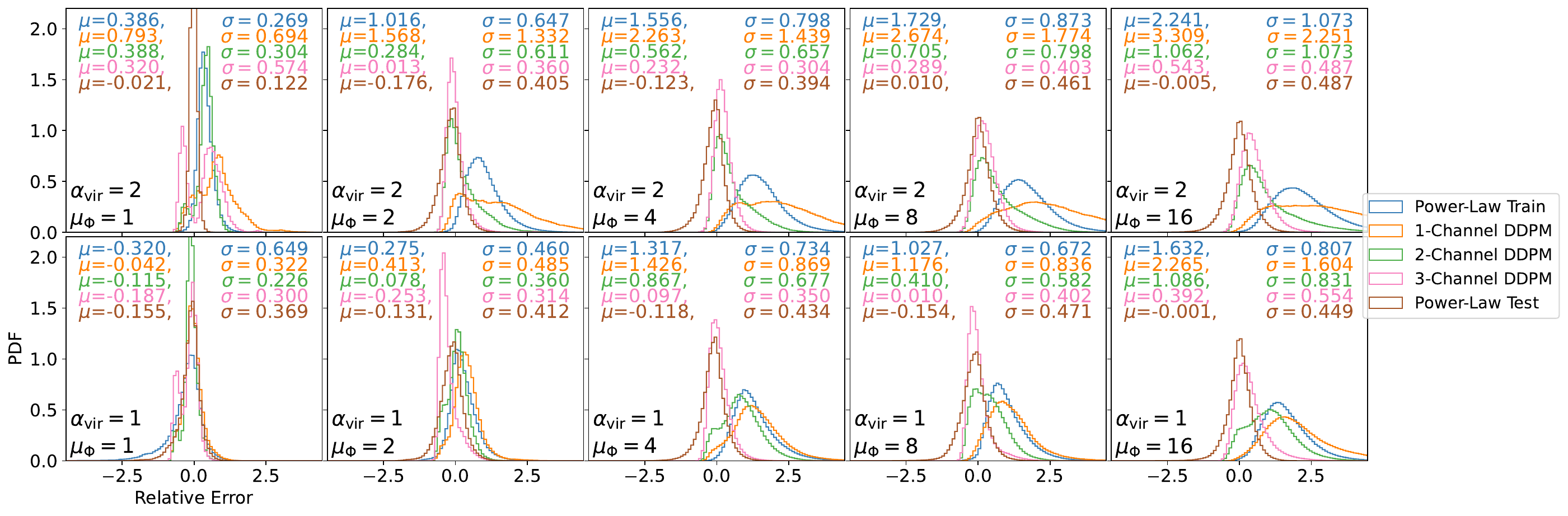}
\caption{Relative error ($\delta_{B}$) distribution between the predicted and true values for the new simulations incorporating self-gravity and outflow feedback, with different virial parameters ($\alpha_{\rm vir}$) and mass-to-flux ratios ($\mu_{\Phi}$).}
\label{fig.ddpm_pred_error_plot_test_Orion_outflow_2p5}
\end{figure*} 

\begin{table*}[!htb]
\begin{center}
\caption{Summary of Relative Error ($\delta_{B}$) for Different Models on New Simulations Including Self-gravity$^a$ \label{tab-model-orion-outflow-ddpm}}
    \begin{tabular}{cccccccccccccc}
    \hline
    \hline
        $\alpha_{\rm vir}$ & $\mu_{\Phi}$ & $B$ & $M_{A}$ & \multicolumn{2}{c}{PL-Train} & \multicolumn{2}{c}{1-Channel} & \multicolumn{2}{c}{2-Channel} & \multicolumn{2}{c}{3-Channel} &  \multicolumn{2}{c}{PL-Test}   \\ 
          &   &  ($\mu$G) &  & mean & std & mean & std & mean & std & mean & std & mean & std   \\ \hline 
         \multirow{5}{*}{2} & 1 & 25.5 & 0.87 & 0.386 & 0.269 & 0.793 & 0.694 & 0.388 & 0.304 & 0.320 & 0.574 & -0.021 & 0.122 \\ 
        ~ & 2 & 12.8 & 1.75 & 1.016 & 0.647 & 1.568 & 1.332 & 0.284 & 0.611 & 0.013 & 0.360 & -0.176 & 0.405 \\ 
        ~ & 4 & 6.4 & 3.50 & 1.556 & 0.798 & 2.263 & 1.439 & 0.562 & 0.657 & 0.232 & 0.304 & -0.123 & 0.394 \\ 
        ~ & 8 & 3.2 & 6.99 & 1.729 & 0.873 & 2.674 & 1.774 & 0.705 & 0.798 & 0.289 & 0.403 & 0.010 & 0.461 \\ 
        ~ & 16 & 1.6 & 13.98 & 2.241 & 1.073 & 3.309 & 2.251 & 1.062 & 1.073 & 0.543 & 0.487 & -0.005 & 0.487 \\ \hline
        \multirow{5}{*}{1} & 1 & 51.1 & 0.62 & -0.320 & 0.649 & -0.042 & 0.322 & -0.115 & 0.226 & -0.187 & 0.300 & -0.155 & 0.369 \\ 
        ~ & 2 & 25.5 & 1.24 & 0.275 & 0.460 & 0.413 & 0.485 & 0.078 & 0.360 & -0.253 & 0.314 & -0.131 & 0.412 \\ 
        ~ & 4 & 12.8 & 2.47 & 1.317 & 0.734 & 1.426 & 0.869 & 0.867 & 0.677 & 0.097 & 0.350 & -0.118 & 0.434 \\ 
        ~ & 8 & 6.4 & 4.94 & 1.027 & 0.672 & 1.176 & 0.836 & 0.410 & 0.582 & 0.010 & 0.402 & -0.154 & 0.471 \\ 
        ~ & 16 & 3.2 & 9.89 & 1.632 & 0.807 & 2.265 & 1.604 & 1.086 & 0.831 & 0.392 & 0.554 & -0.001 & 0.449 \\ \hline
   \multicolumn{14}{p{0.5\linewidth}}{Notes:}\\
\multicolumn{14}{p{0.83\linewidth}}{$^a$ Virial parameter, mass-to-flux ratios, Alfv\'en mach number, and the mean and standard deviation of the relative error ($\delta_{B}$) for predictions from various models. From left to right: power-law fitting from the training set, 1-channel DDPM, 2-channel DDPM, 3-channel DDPM, and power-law fitting based on the column density from the corresponding new simulations including self-gravity (test set).}
\end{tabular}
\end{center}
\end{table*}

In this section, we evaluate the performance of the DDPMs on new simulations generated with a different code and under varying physical and initial conditions. These new MHD simulations follow the setup in \citet{2023ApJ...942...95X}. We run ideal MHD simulations using the \orion\ code \citep{2021JOSS....6.3771L} to model turbulent clouds with periodic boundary conditions, excluding self-gravity. The simulation box size is $5\times5\times5$~pc$^3$, with the magnetic field initialized along the $z$-axis. Turbulence is driven with equal energy distribution between solenoidal and compressive modes. {The turbulence driving occurs on large scales, specifically in Fourier space at wavenumbers corresponding to 1/2 - 1 of the box size, with an appropriate decay time to maintain driving mode correlations for about two crossing times.} The gas is modeled as an isothermal ideal gas with a temperature of 10~K. The 3D Mach number is 10.5, positioning the simulated cloud on the linewidth-size relation, $\sigma_{\rm 1D}=0.72 R^{0.5}_{\rm pc}$ \kms \citep{2007ARA&A..45..565M}. The base grid for these calculations is 256$^3$, without adaptive mesh refinement (AMR).

Simulations are performed with two different virial parameters, $\alpha_{\rm vir}=5\sigma_v^2 R/(GM)=$ 1 and 2. Additionally, five different mass-to-flux ratios are adopted: $\mu_{\Phi}=M_{\rm gas}/M_{\rm \Phi}=2\pi G^{1/2}M_{\rm gas}/(BL^2)$, with $\mu_{\Phi}$= 1, 2, 4, 8, and 16. This results in 10 different simulation setups, with Alfv\'en Mach numbers ranging between 0.62 and 14, and initial magnetic field strengths ranging from 1.6~$\mu$G to 51~$\mu$G.

Figures~\ref{fig.ddpm_pred_hist_channel_Orion_alpha2} and \ref{fig.ddpm_pred_hist_channel_Orion_alpha1} display 2D histograms comparing the ground truth magnetic field strength with the inferred values from various models on the new simulations, categorized by virial parameters and mass-to-flux ratios. Figure~\ref{fig.ddpm_pred_error_plot_test_Orion} provides a quantitative analysis of the relative error ($\delta_{B}$) distributions across these simulations, with a summary of the results in Table~\ref{tab-model-orion-ddpm}. The results show that under different initial conditions, the power-law fitting inferred from the training set often leads to systematic offsets due to variations in the $B-N$ relation. Similarly, the 1-channel and 2-channel DDPMs exhibit notable prediction offsets on the new simulation data. In contrast, the 3-channel DDPM consistently performs better than the other models across different conditions.

Upon examining the relative error ($\delta_{B}$) performance across different physical conditions, it is noteworthy that all four models—including the power-law fitting from the training set and the three DDPMs—achieve nearly the same accuracy as the power-law fitting based on the new simulations when $\alpha_{\rm vir}=1$ and $\mu_{\Phi}=1$. This suggests that such conditions are well-represented in the training data. Conversely, when the relative error ($\delta_{B}$) of the predicted magnetic field strength from the training set's power-law fitting deviates significantly from zero, it indicates that these conditions—such as those with $\mu_{\Phi}>4$—are underrepresented in the training set.

In these outlier cases, both the 1-channel and 2-channel DDPMs exhibit performance similar to the power-law fitting from the training set, showing significant deviations from the true values. In contrast, the 3-channel DDPM delivers superior results, with much smaller relative error ($\delta_{B}$) offsets. This demonstrates the 3-channel DDPM's ability to effectively handle out-of-distribution data, a critical skill for applying the model to real observational datasets, which often contain previously unseen conditions for machine learning models.

To further demonstrate the ability of the 3-channel DDPM method to learn from out-of-distribution data, we tested the models on turbulent box simulations that include both self-gravity and outflow feedback mechanisms. Each simulation in these datasets features at least one protostellar outflow. {It is important to note that outflows can partially offset the rapid decay of turbulent energy. However, the impact of numerical dissipation in previous simulations, particularly those lacking feedback mechanisms (e.g., in turbulence-dominated diffuse regions), remains uncertain.} Figures~\ref{fig.ddpm_pred_hist_channel_Orion_outflow_2p5_alpha2} and \ref{fig.ddpm_pred_hist_channel_Orion_outflow_2p5_alpha1} show 2D histograms comparing the ground truth magnetic field strength with the inferred values from different models applied to these new simulations, taking into account self-gravity and outflow feedback. The simulations are categorized by virial parameters and mass-to-flux ratios. Figure~\ref{fig.ddpm_pred_error_plot_test_Orion_outflow_2p5} provides a quantitative analysis of the relative error ($\delta_{B}$) distributions across these simulations, with a summary of results in Table~\ref{tab-model-orion-outflow-ddpm}. It is clear that the 3-channel DDPM consistently outperforms the other models under various physical conditions, demonstrating its significantly improved accuracy in predicting magnetic field strength on previously unseen data.


\subsection{Comparison with the DCF Method}
\label{Comparison with the DCF Method}

\begin{figure*}[hbt!]
\centering
\includegraphics[width=0.99\linewidth]{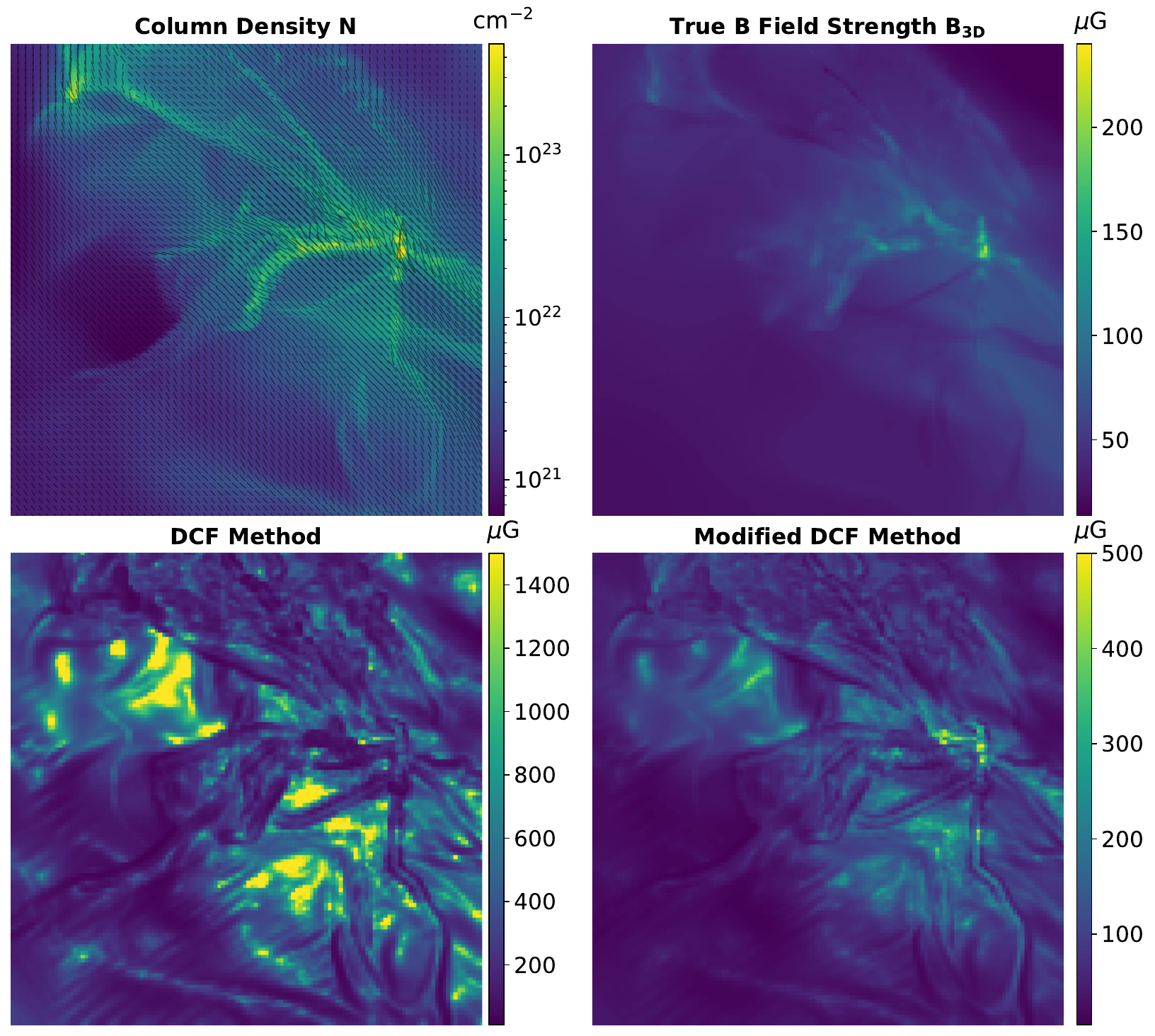}
\caption{An example illustrating the magnetic field strength calculated using the DCF method: column density map with magnetic field directions (top left), true magnetic field strength (top right), classical DCF method result (bottom left), and modified DCF method result (bottom right).}
\label{fig.simulation_DCF_img}
\end{figure*} 

\begin{figure}[hbt!]
\centering
\includegraphics[width=0.99\linewidth]{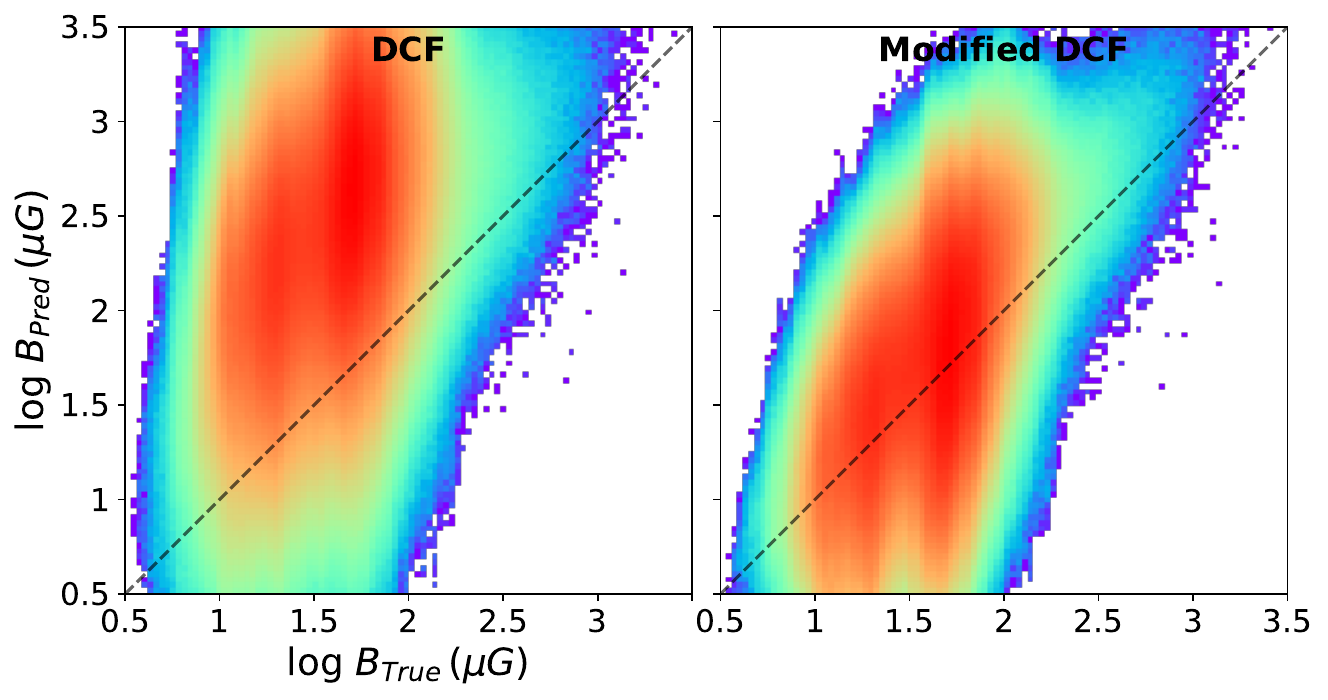}
\caption{2D histograms comparing the ground truth magnetic field strength with inferred values using the DCF methods: classical DCF method (left panel) and modified DCF method (right panel).}
\label{fig.simulation_DCF_hist_img}
\end{figure} 

\begin{figure}[hbt!]
\centering
\includegraphics[width=0.99\linewidth]{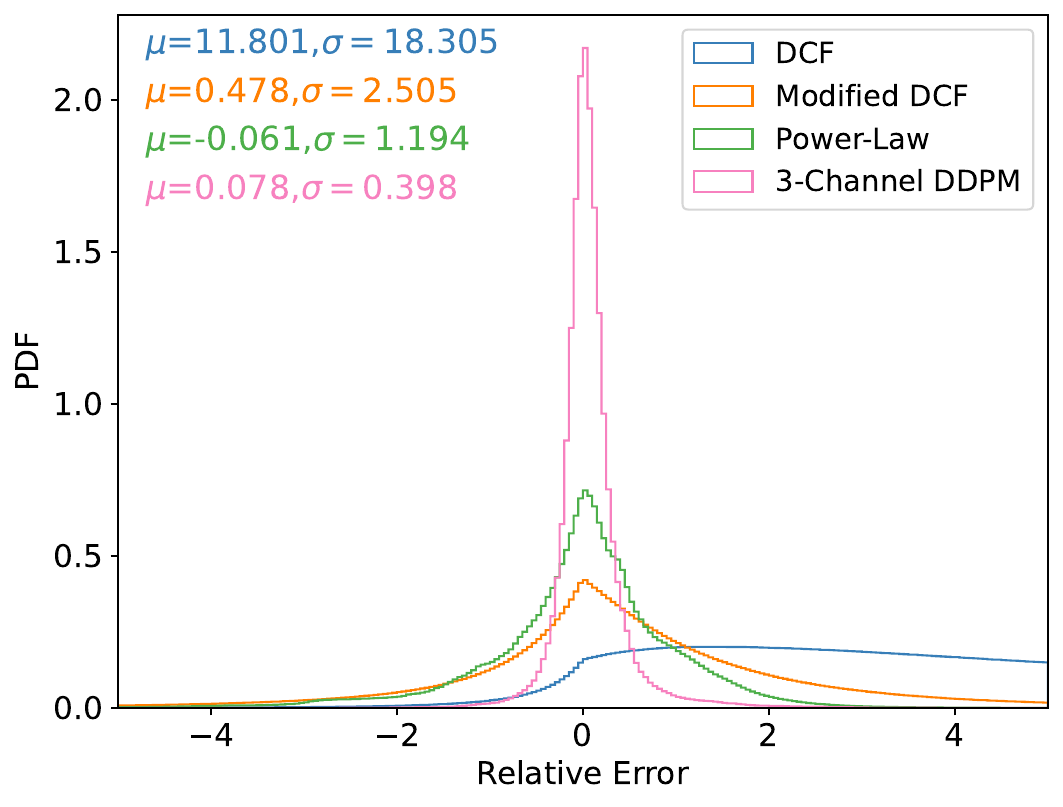}
\caption{Relative error ($\delta_{B}$) distribution between the predicted and true values for various methods, including the classical and modified DCF methods. For comparison, the DDPM and power-law fitting results are also included. }
\label{fig.ddpm_pred_error_plot_DCF}
\end{figure} 

In this section, we assess the performance of the DCF method in calculating magnetic field strength for the simulations discussed in Section~\ref{Magnetohydrodynamics Simulations} and compare the results with machine learning approaches. We follow the approach described in \citet{2024ApJ...967..157L}, which employs two variations of the DCF method. The first is the classical DCF method for estimating the POS magnetic field strength, as outlined in Equation~\ref{eqn_dcf}. The second is the modified DCF method, as described in Equation~\ref{eqn_modified_dcf}. We applied a $2\times2$ sliding window to calculate the polarization angle dispersion $\sigma_{PA}$ at the pixel level. The POS magnetic field strength was then calculated for each pixel using Equations~\ref{eqn_dcf} and \ref{eqn_modified_dcf}. Assuming the three components of the 3D magnetic field have comparable magnitudes, we estimated the total magnetic field strength as $B_{\text{3D}} = \sqrt{3} B_{\text{1D}} = \frac{\sqrt{3}}{\sqrt{2}} B_{\text{POS}}$.

Figure~\ref{fig.simulation_DCF_img} provides an example of the magnetic field strength calculated using both the classical and modified DCF methods. Figure~\ref{fig.simulation_DCF_hist_img} shows 2D histograms comparing the ground truth magnetic field strength with the inferred values from the DCF methods. Figure~\ref{fig.ddpm_pred_error_plot_DCF} presents the relative error ($\delta_{B}$) distributions for various methods, including the classical and modified DCF approaches.

It is evident that the classical DCF method overestimates the magnetic field strength by about an order of magnitude, aligning with the results reported by \citet{2021A&A...647A.186S}, and shows a large dispersion in $\delta_{B}$ of 18. The modified DCF method performs significantly better, with an average overestimation of 48\% and a reduced dispersion of 2.5. However, while the modified DCF method shows improvement over the classical version, it still falls short of the accuracy achieved by the power-law fitting approach and the 3-channel DDPM. 

To assess the impact of window size on the DCF method, we employ varying window sizes to scan the image and calculate the density-weighted ground truth magnetic field strength within each window. Additionally, we compute the density-weighted LOS velocity dispersion, angle dispersion, and mean density within the selected windows to estimate the magnetic field strength using the DCF method, as detailed in Appendix~\ref{Impact of Window Size on DCF}. Our analysis reveals that the choice of window size significantly affects the performance of the DCF method. However, there is no universally optimal window size, even at the image scale. For instance, when applying a $128\times128$ window, the DCF method does not demonstrate improved accuracy in estimating the average magnetic field strength on larger scales. This underscores the inherent limitations of the DCF method in achieving consistent performance across different window sizes.

There are several reasons why the DCF method performs poorly in simulations. First and foremost, the fundamental assumption of the DCF method--that turbulent energy and magnetic energy are in equipartition--does not always hold in molecular clouds. Additionally, the presence of gravity further disrupts this energy equipartition assumption. Furthermore, from a technical perspective, the LOS velocity dispersion in our simulations is influenced not only by turbulence but also by large-scale motions, e.g., cloud collisions, which can lead to a significant overestimation of the magnetic field strength \citep{2023MNRAS.524.4431H}. Moreover, since the DCF method is inherently a statistical approach, calculating the polarization angle dispersion on a small scale (e.g., using a $2\times2$ sliding window) introduces considerable uncertainty. Consequently, caution should always be exercised when estimating magnetic field strength using DCF methods.


\section{Conclusions}
\label{Conclusions}

We trained the deep learning model DDPM to predict magnetic field strength from observables, including column density, polarization angles, and LOS velocity dispersion. We evaluated the performance of the diffusion model on both synthetic test samples and new simulation data that are outside the distribution of the training data. Our main findings are summarized below:

\begin{enumerate}

\item  There is a power-law correlation between magnetic field strength and column density; however, the power-law exponents vary with different initial magnetic field strengths and dynamic conditions. This variability makes it challenging to accurately infer magnetic field strength based solely on column density across different datasets.

\item  We trained three DDPMs: the 1-channel DDPM (using column density as the only input), the 2-channel DDPM (using both column density and polarization angle), and the 3-channel DDPM (using column density, polarization angle, and LOS nonthermal velocity dispersion). The 3-channel DDPM consistently outperformed both the other DDPM models and the power-law fitting approach based solely on column density from the training set.

\item  We also tested the DDPMs on new simulations generated using a different code and under different physical and initial conditions. The 3-channel DDPM showed the best performance, with the smallest systematic offset in the relative error. This suggests that the 3-channel DDPM is highly capable of handling unseen data that were not part of the training set.

\item  Additionally, we compared the DCF methods (both the classical and modified versions) with the DDPM predictions for estimating magnetic field strength in the simulations. The classical DCF method overestimated the magnetic field strength by about an order of magnitude. While the modified DCF method improved on the classic version, it still fell short of the precision achieved by the 3-channel DDPM.

\end{enumerate}

In a forthcoming companion paper, we will apply the 3-channel DDPM trained in this study to real observational datasets of molecular clouds, including from the Polarized Light from Massive Protoclusters (POLIMAP) survey \citep{2024ApJ...967..157L}, which has used SOFIA-HAWC+ to map the polarized dust continuum emission of a sample of Infrared Dark Clouds (IRDCs), to systematically measure the magnetic field strength in these regions via DCF-type methods. 

{We express our gratitude to the anonymous referee for their valuable comments and suggestions, which have significantly enhanced the quality of this paper.} D.X. acknowledges support from the Virginia Initiative on Cosmic Origins (VICO). D.X. acknowledges the support of the Natural Sciences and Engineering Research Council of Canada (NSERC), [funding reference number 568580]. D.X. also acknowledges support from the Eric and Wendy Schmidt AI in Science Postdoctoral Fellowship Program, a program of Schmidt Sciences. J.C.T. acknowledges support from NSF grant AST-2009674 and ERC Advanced grant MSTAR. J.K. acknowledges funding from the Chalmers Astrophysics and Space Sciences Summer (CASSUM) program. The authors acknowledge Research Computing at The University of Virginia for providing computational resources and technical support that have contributed to the results reported within this publication. The authors further acknowledge the use of NASA High-End Computing (HEC) resources through the NASA Advanced Supercomputing (NAS) division at Ames Research Center to support this work.

\appendix

\section{Impact of Window Size on DCF}
\label{Impact of Window Size on DCF}

\begin{figure*}[hbt!]
\centering
\includegraphics[width=0.49\linewidth]{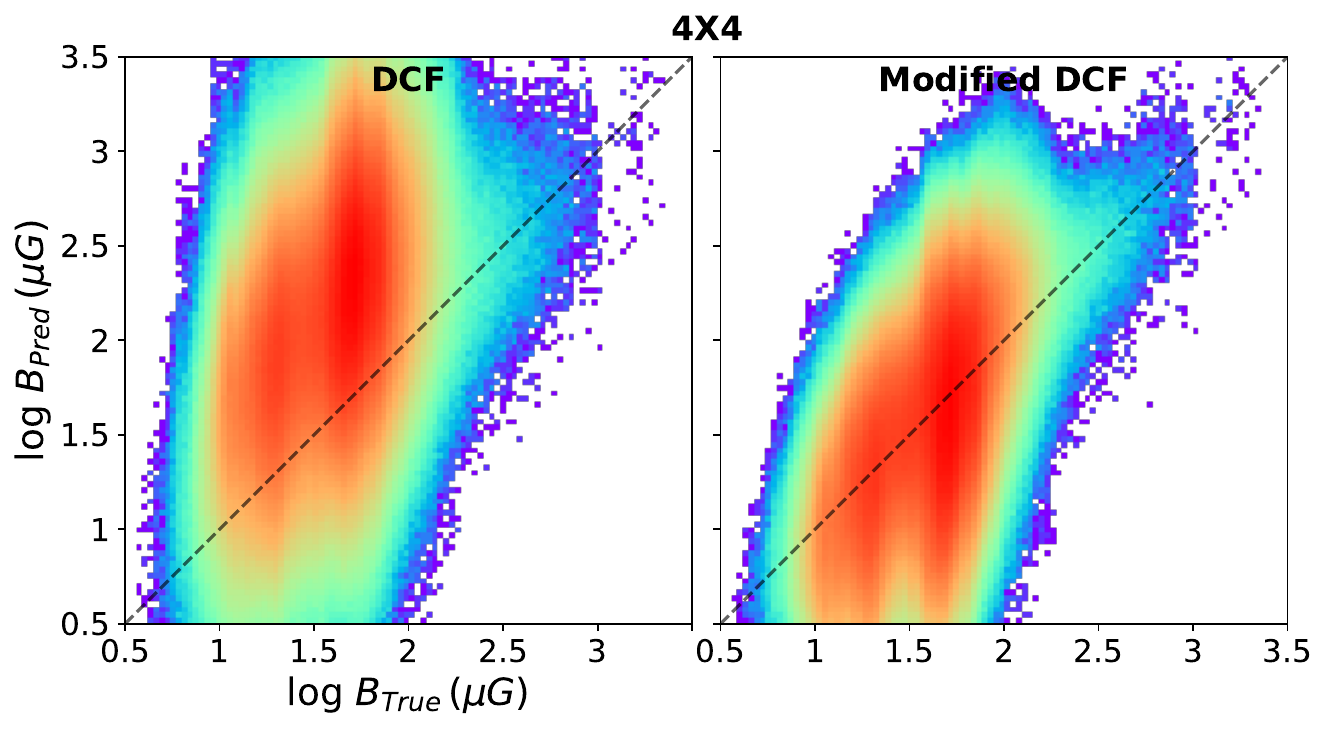}
\includegraphics[width=0.49\linewidth]{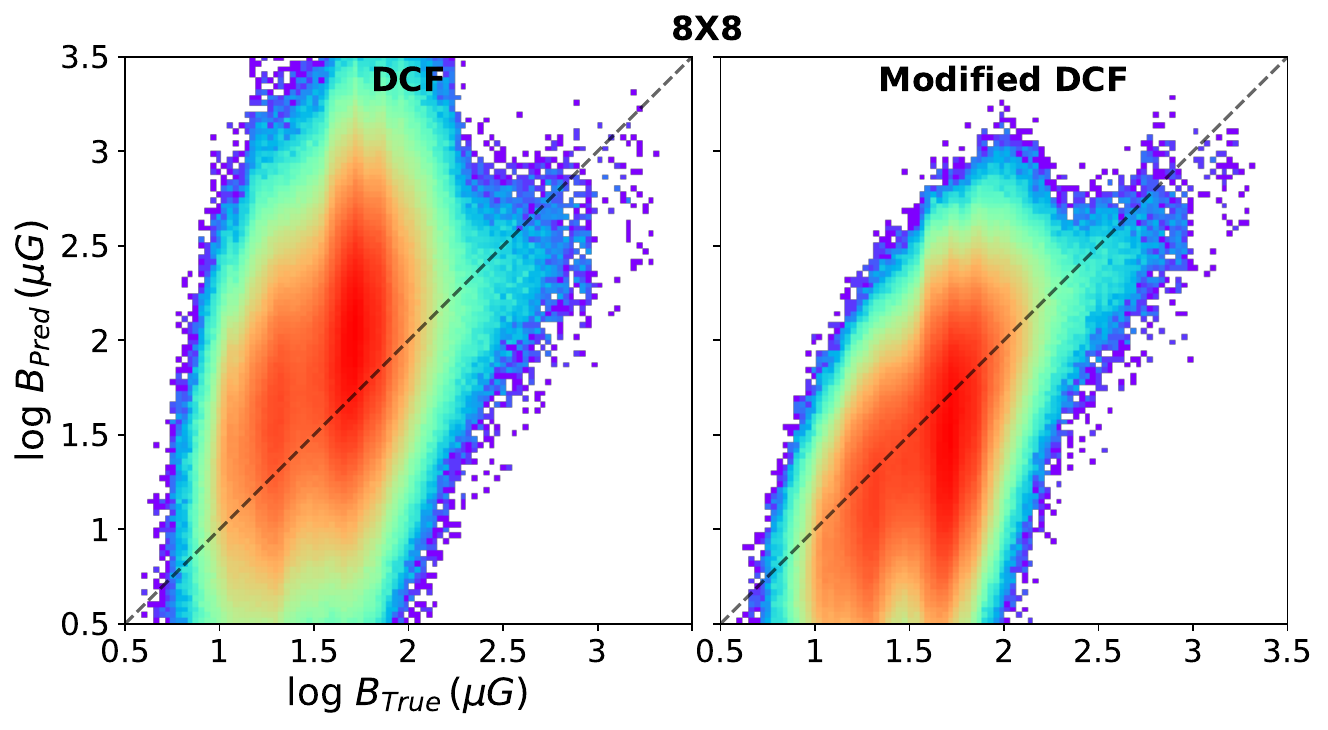}
\includegraphics[width=0.49\linewidth]{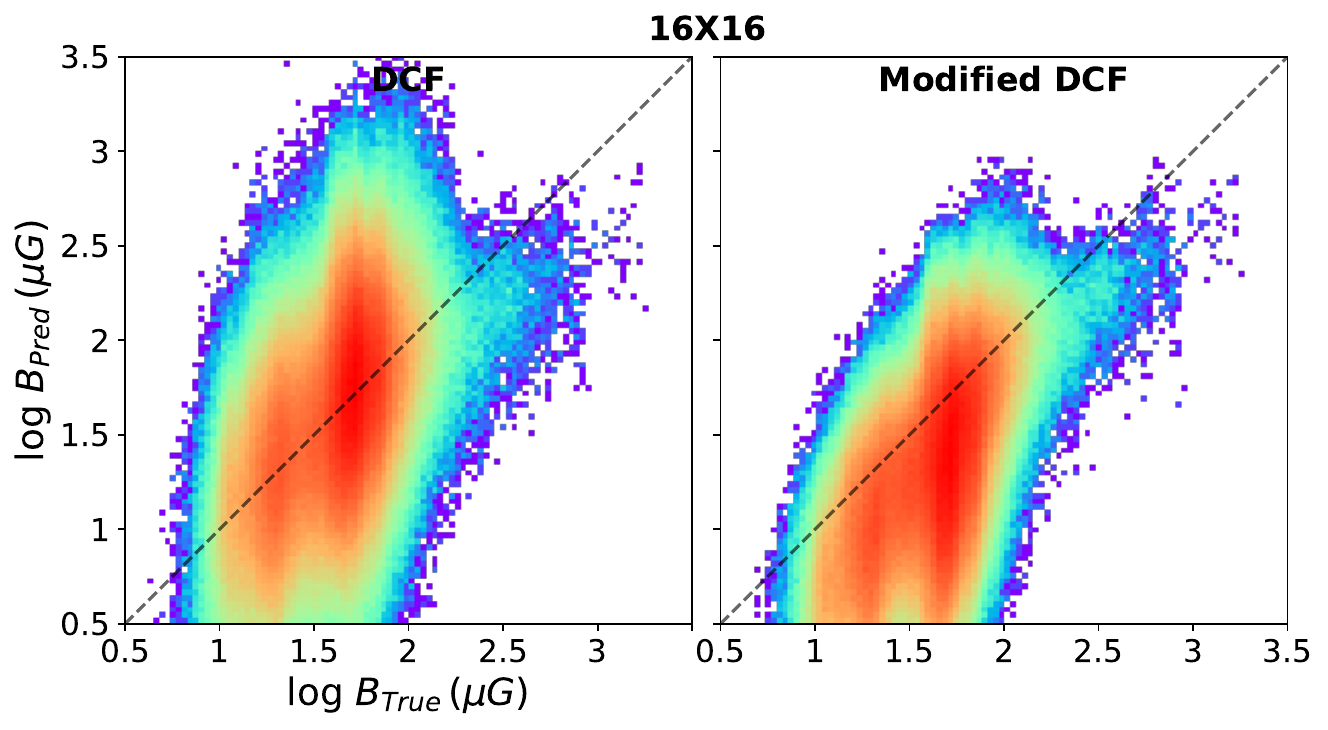}
\includegraphics[width=0.49\linewidth]{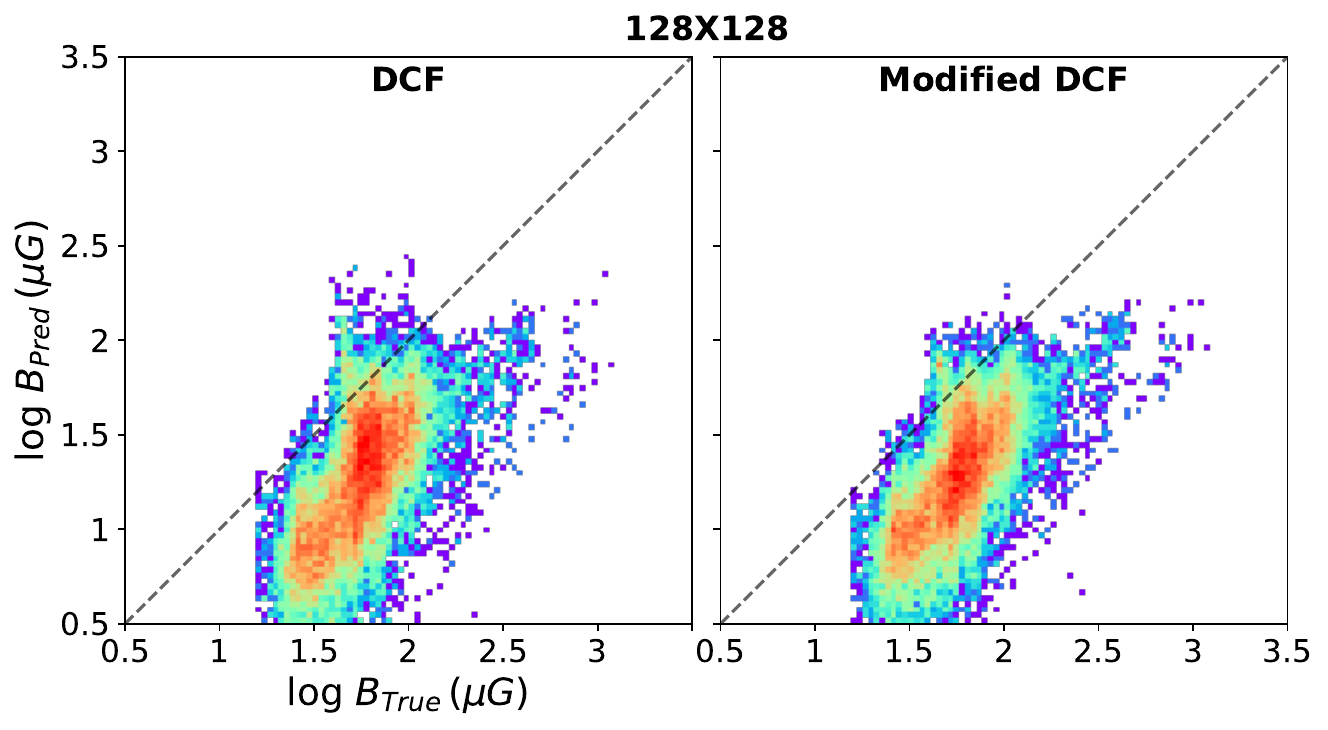}
\caption{2D histograms comparing the ground truth magnetic field strength with inferred values using the DCF methods with different window sizes: classical DCF method (left panel) and modified DCF method (right panel).}
\label{fig.simulation_DCF_hist_img_window}
\end{figure*}

\begin{figure}[hbt!]
\centering
\includegraphics[width=0.48\linewidth]{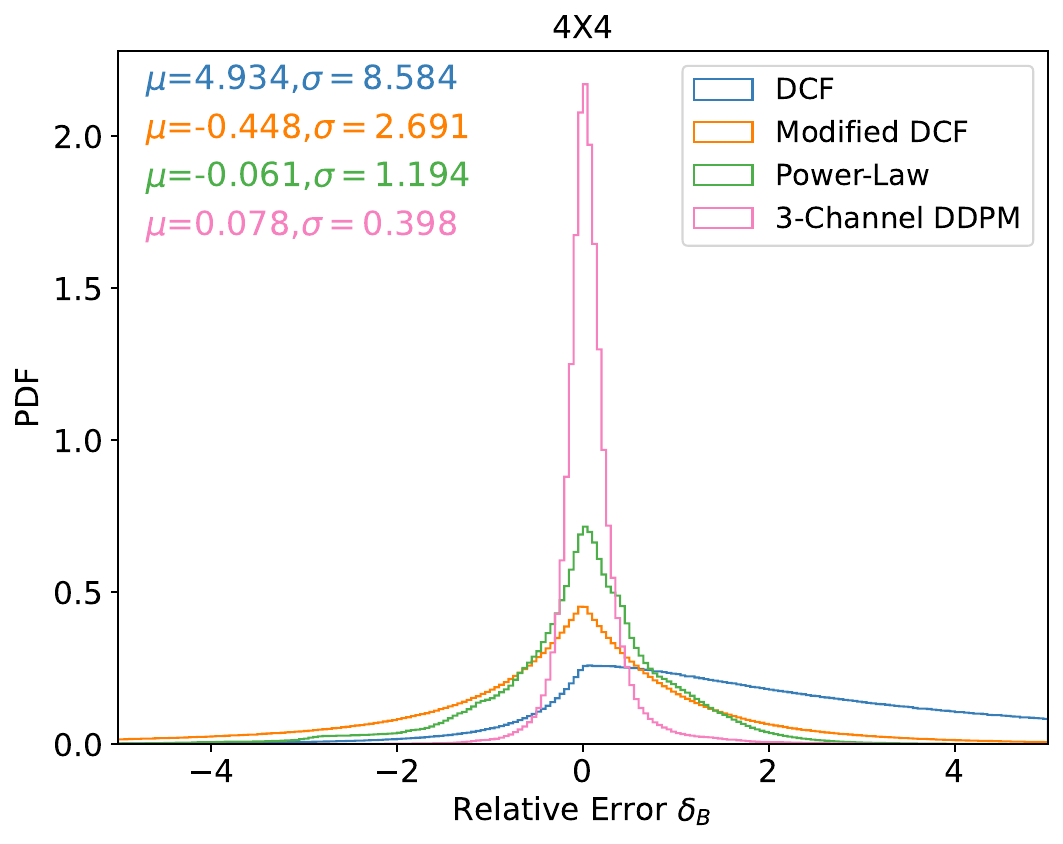}
\includegraphics[width=0.48\linewidth]{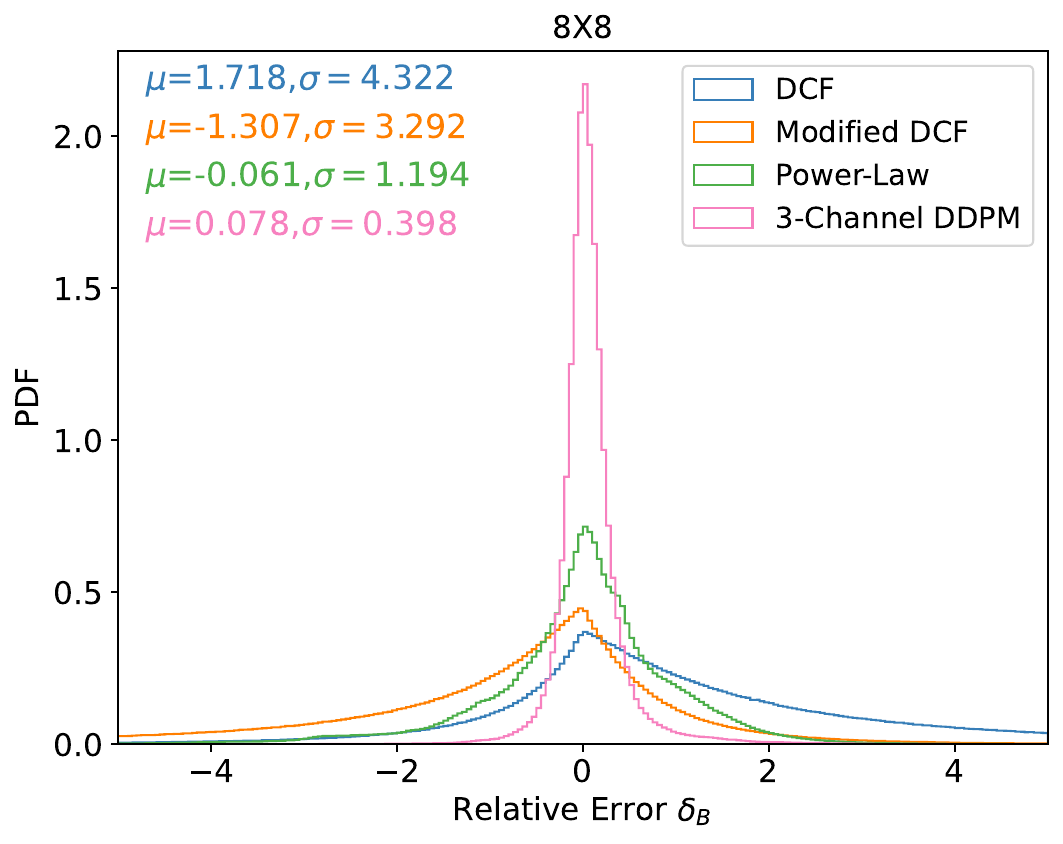}
\includegraphics[width=0.48\linewidth]{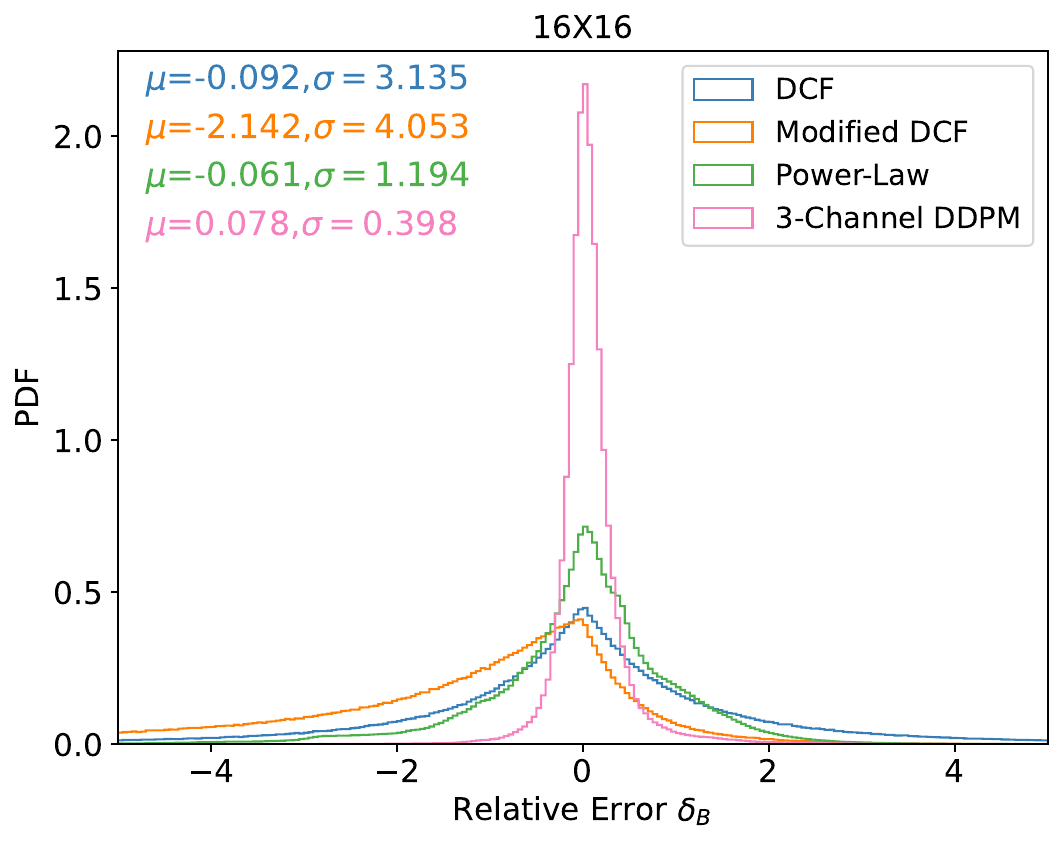}
\includegraphics[width=0.48\linewidth]{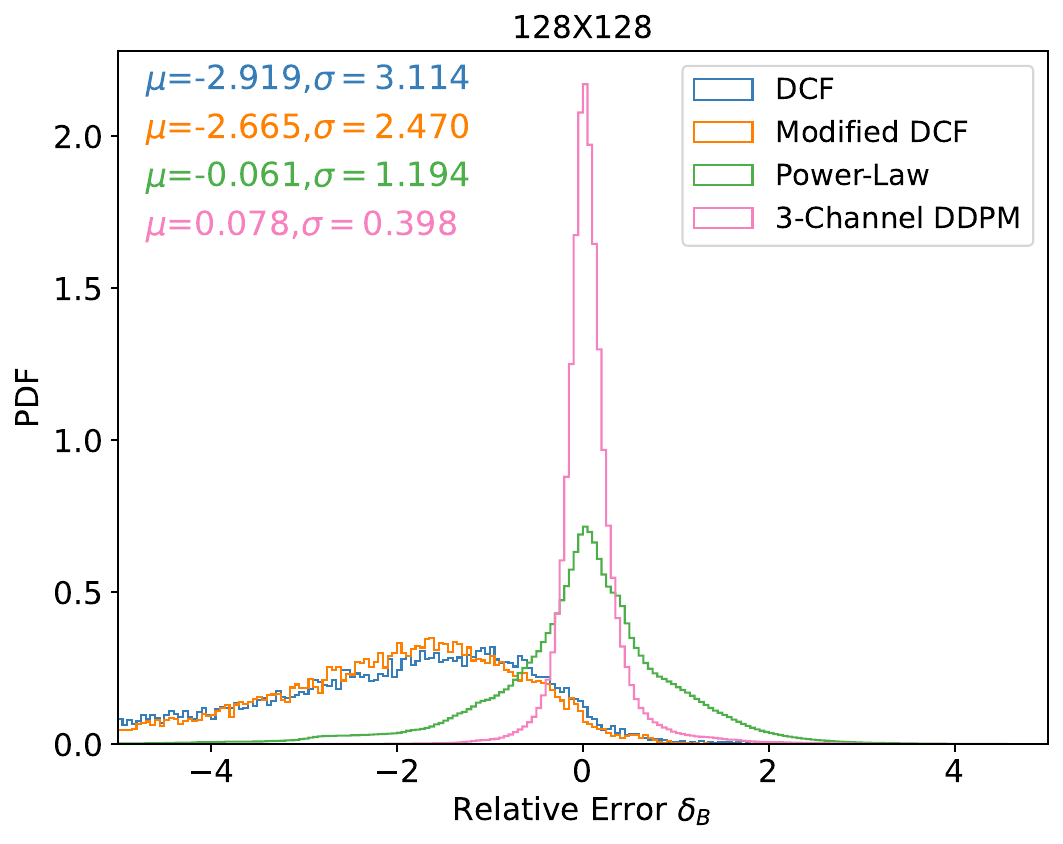}
\caption{Relative error ($\delta_{B}$) distribution between the predicted and true values for various methods, including the classical and modified DCF methods with different window sizes. For comparison, the DDPM and power-law fitting results are also included.}
\label{fig.ddpm_pred_error_plot_DCF_window}
\end{figure}

In this section, we evaluate the effect of window size on the performance of the DCF method by utilizing various window sizes to scan the image. For each window, we calculate the density-weighted ground truth magnetic field strength and the corresponding density-weighted LOS velocity dispersion, angle dispersion, and mean density to estimate the magnetic field strength using the DCF method. The window sizes tested include $4\times4$, $8\times8$, $16\times16$, and $128\times128$. Figures~\ref{fig.simulation_DCF_hist_img_window} and~\ref{fig.ddpm_pred_error_plot_DCF_window} illustrate the comparisons between the ground truth magnetic field strengths and those estimated by the DCF method for these different window sizes. The results indicate that there is no universally optimal window size, even at the image scale. For instance, using a $128 \times 128$ window does not lead to improved accuracy in estimating the average magnetic field strength on larger scales. This result differs from previous studies, such as \citet{2022MNRAS.510.6085L}, where the DCF method demonstrated strong performance in simulations of filament formation. The discrepancy may arise from the fact that our simulations encompass a much broader range of physical conditions, where the fundamental assumptions of the DCF method may no longer hold.

\bibliographystyle{aasjournal}
\bibliography{references}

\end{CJK*}

\end{document}